\documentclass[11pt,twoside]{article}
\usepackage{asp2010_hinode5}
\newcommand\at{\makeatletter@\makeatother }

\resetcounters

\begin{document}

\title{Recent Advances in the Exploration of the Small-Scale Structure of the 
Quiet Solar Atmosphere: Vortex Flows, the Horizontal Magnetic Field, and the 
Stokes-{\boldmath $V$} Line-Ratio Method}
\author{O.~Steiner and R.~Rezaei,
\affil{Kiepenheuer-Institut f\"ur Sonnenphysik, Sch\"oneckstrasse 6,
       79104 Freiburg, Germany}
}

\begin{abstract}
We review (i) observations and numerical simulations of vortical flows in the solar
atmosphere  and (ii) measurements of the horizontal magnetic field in quiet Sun 
regions. First, we discuss various manifestations of vortical flows and 
emphasize the role of magnetic fields in mediating swirling motion created near the 
solar surface to the higher layers of the photosphere and to the chromosphere. 
We reexamine existing simulation runs of solar surface magnetoconvection with regard
to vortical flows and compare to previously obtained results.
Second, we review contradictory results and problems 
associated with measuring the angular distribution of the magnetic field in quiet
Sun regions. Furthermore, we review the Stokes-$V$-amplitude ratio method for
the lines \ion{Fe}{i} $\lambda\lambda$ 630.15 and 630.25~nm. We come to the conclusion
that the recently discovered two distinct populations in scatter plots of this ratio
must not bee interpreted in terms of ``uncollapsed'' and ``collapsed'' fields but 
stem from weak granular magnetic fields and weak canopy fields located
at the boundaries between granules and the intergranular space. Based on
new simulation runs, we reaffirm earlier findings of a predominance of the
horizontal field components over the vertical one, particularly in the upper 
photosphere and at the base of the chromosphere.
\end{abstract}

\section{Introduction}
Two different topics are reviewed in this contribution. (1) Vortical flows, which
have been observed to exist in the deep photosphere and in the chromosphere. Over the 
past three years, they have anew attracted the attention from observers and computational 
physicists. We attempt to find and emphasize interrelations 
between the various  manifestations of vortical flows in the solar atmosphere. 
(2) The polarimetry of the horizontal magnetic field in quiet Sun regions. We
review the disparate results that have been obtained in the past and highlight the 
difficulties associated with such measurements. We discuss biases and propose an
alternative method for dealing with selection criteria. The main part of the
second chapter is devoted to the Stokes-$V$ amplitude-ratio method and the origin of
the dichotomy observed in scatter plots of the $V$ amplitude of 
\ion{Fe}{i} 630.15\,nm vs.\ the $V$ amplitude of  \ion{Fe}{i} 630.25\,nm.

\section{Vortical Flows and Vortex Tubes}
\label{sec:vortical_flows}
Vortical flows, swirls, whirlpools, vortex tubes in the solar atmosphere
have become a focus of intense research in the past
three years. This chapter presents a brief review of this research focus
and attempts a comparison of some of these results with simulations carried
out with the CO5BOLD \citep{ost_freytag+al2012} code by the authors. 

Vortical flows in the photosphere
of the Sun were reported long before the present revival. Most notably is
the vortex flow of \citet{ost_brandt+al1988}, which was a vortical movement
of granules that persisted for 1.5~h. These authors conjectured that if such
vortices were a common feature of the solar convection zone, they might
``provide an important mechanism for heating of stellar chromospheres
and coronae by twisting the footpoints of magnetic flux tubes''.

From numerical simulations, \citet{ost_nordlund1985} reports ``a circular motion 
around the center of the downdraft, and the circular velocity is amplified
as the downdraft narrows (`bath-tub' or `inverted tornado')''. The centripetal
force associated with this vortical flow was so strong that the gas pressure
gradient opposing it, apparently led to numerical difficulties because the
gas pressure in the center of the downdraft became too small or negative.
Similar to \citet{ost_brandt+al1988}, he also wondered about the consequences
of such `inverted tornadoes' for the higher layers of the atmosphere and
conjectured that ``the fact that any vertical magnetic field lines in the 
surrounding photosphere must be carried towards, and `sucked into', these 
downdrafts  also makes the phenomenon potentially very important as source 
of hydromagnetic disturbances''.

\citet{ost_wang+al1995} computed from a time series of continuum images from the
Pic du Midi Observatory the horizontal vector flow field and associated
divergence and the vertical component of the curl and found that the curl is 
correlated with regions of negative divergence, suggesting excess vorticity 
in intergranular lanes. A similar result is reported by \citet{ost_poetzi+brandt2007}.

\subsection{Vortical Flows in the Photosphere}
\label{sec:bonet_swirls}
With the ever increasing spatial resolution and homogeneity of long duration
image sequences thanks to larger apertures of ground based solar telescopes,
to high order adaptive optics systems and advanced image restoration 
techniques, and thanks to space based and balloon borne telescopes, vortical 
flows on ever smaller scales became visible.

\begin{table}[!ht]
\caption{Vortex properties as detected by \citet{ost_bonet+al2008,ost_bonet+al2010}. 
The numbers in parentheses in the second 
column refer to the number of vortices with clockwise (first) and counterclockwise 
(second) sense of rotation.\label{tab:bonet}}
\smallskip
\begin{center}
{\small
\begin{tabular}{crlcc}
\tableline
\noalign{\smallskip}
 telescope & \multicolumn{2}{c}{number of detected} & space-time density         & mean life-time \\[0ex]
           & \multicolumn{2}{c}{vortices [-]}       & [Mm$^{-2}$\,minute$^{-1}$] & [minutes]      \\
\noalign{\smallskip}
\tableline
\noalign{\smallskip}
SST     & 138 &(68, 70)       & $1.8\times 10^{-3}$        & 5.1             \\
Sunrise & 42  &(15, 27)       & $3.1\times 10^{-3}$        & 7.9             \\
\noalign{\smallskip}
\tableline
\end{tabular}
}
\end{center}
\end{table}

Thus, \citet{ost_bonet+al2008} discovered small whirlpools on the solar surface 
with a size similar to terrestrial hurricanes ($\la 0.5$~Mm), using the
Swedish Solar Telescope (SST). They detected
them because some magnetic bright points follow a logarithmic spiral on their 
way to being engulfed by a downdraft. Their image sequences at disk-center 
show $0.9\times 10^{-2}$ vortices per square Mm, with a lifetime of the order 
of 5~minutes, and with no preferred sense of rotation (see Table~\ref{tab:bonet}).
They repeated this kind of analysis with data from the ballon borne solar
telescope Sunrise \citep{ost_barthol2010} and obtained similar results 
(see Table~\ref{tab:bonet}) with
the notable difference that from the Sunrise data, they detected a preferred 
sense of rotation of the vortices. They speculate that this might be due to 
the fact that the Sunrise image sequences were recorded away from the
solar equator, different from the previously obtained SST data. Away from
the equator, differential rotation may have introduced a preferred sense of
rotation. It remains to be explored if this effect is indeed taking place.

\citet{ost_vargas-dominguez+al2011} study a region of $69\arcsec\times 69\arcsec$ 
of quiet Sun that includes the field of view investigated by \citet{ost_bonet+al2008}
and derive statistical properties of swirl motions in the photosphere. They
confirm the value for the space-time density of \citet{ost_bonet+al2008} and
find no significant preference in the sense of rotation.

On a slightly larger scale, \citet{ost_attie+al2009} identify two long lasting vortex 
flows located at supergranular junctions, based on two time series of granulation 
from the Hinode/SOT broad band imager. The first vortex flow lasts at least 1~h 
and is $\approx 20\arcsec$ wide, the second vortex flow lasts more than 2~h and is 
$\approx 27\arcsec$ wide. In one case, the corresponding magnetogram  shows a
magnetic element of polarity opposite to the magnetic field in the center of the 
vortex, moving around the vortex center. At the same time, there is enhanced 
\ion{Ca}{ii} emission at the vortex center, which suggests possible magnetic
dissipation because of entwinement of the magnetic fields.

\citet{ost_balmaceda+al2010} report an observation of magnetic flux concentrations being 
dragged towards the center of a convective vortex motion in the solar photosphere
from data, including magnetogram sequences, simultaneously acquired with the SST 
and the Solar Optical Telescope (SOT) of Hinode. They argue that these small-scale 
motions were ``likely to play a role in heating the upper solar atmosphere by 
twisting magnetic flux tubes.'' Likewise, \citet{ost_manso-sainz+al2011} find two
instances of magnetic elements which describe a vortical motion on a scale of 
$\la 400$~km, using spectro-polarimetric and magnetographic data obtained with Hinode.

\subsection{Swirls in the Chromosphere}
\label{sec:wedemeyer_swirls}
\begin{figure}[t]
\centerline{\includegraphics[width=1.0\textwidth]{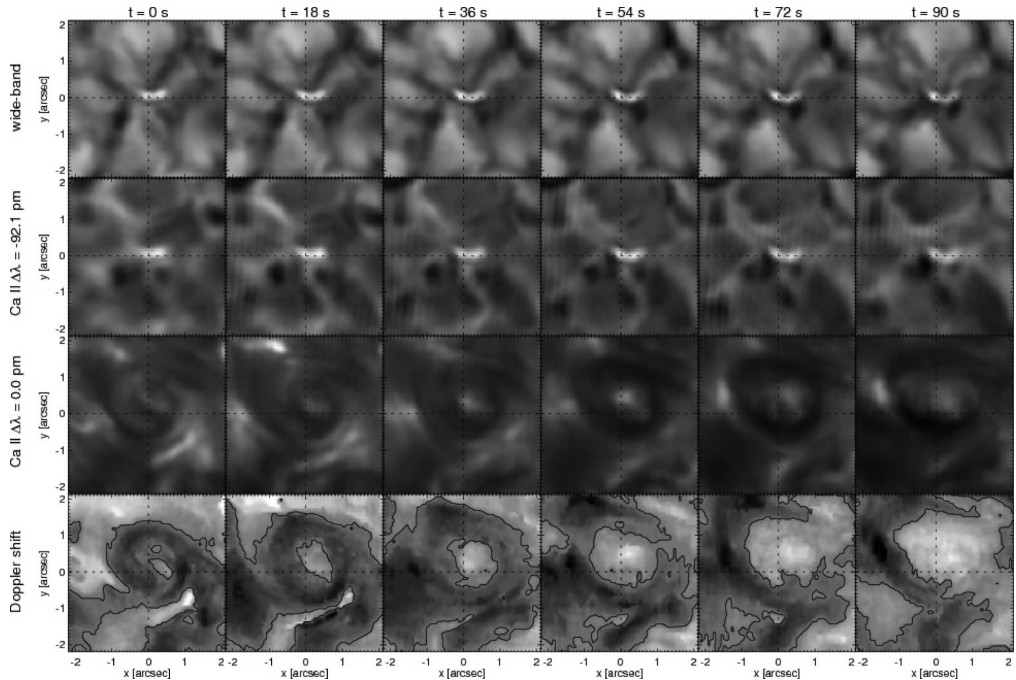}}
\caption{Temporal evolution of a swirl event as seen in close-ups of intensity 
maps in the wide-band (top row), Ca line wing (upper middle), Ca line
core (lower middle), and Doppler shift (bottom). The black contours in
the bottom row mark zero Doppler shift. The grey scale of the Doppler shift 
is from $-5.5$ to $+5.5$~km\,s$^{-1}$ with negative values corresponding to
blueshifts and thus upflows. Credit: \citet{ost_wedemeyer+rouppe2009} reproduced 
with permission \copyright\ ESO.
\label{fig:swirl_wedemeyer}}
\end{figure}

Using the Swedish Solar Telescope (SST) with the CRISP Fabry P\'erot system, 
\citet{ost_wedemeyer+rouppe2009} found in narrow-band filtergrams with a FWHF of 
11.1~pm in the line core of \ion{Ca}{i} 854.2~nm swirling motions in 
chromospheric layers of a quiet Sun region inside a coronal hole. There,
the view is not obstructed by chromospheric filaments, which gives free sight 
to chromospheric layers that would not be easily accessible in more active regions
(see Figure~\ref{fig:swirl_wedemeyer}).

The size of the swirl events is comparable to large terrestrial typhoons
but their physical origin is a very different one. This is obvious as co-temporal
broad band images show close groups of photospheric bright points that move 
with respect to each other, right beneath the swirls. Since such bright points 
are a reliable signature of tiny magnetic flux concentrations in the photosphere,
it is highly probable that the origin of the chromospheric swirls are to be found in 
these moving magnetic flux concentrations. The swirls also exhibit Doppler shifts of 
$-2$ to $-4$~km\,s$^{-1}$ with peak values of up to $-7$~km\,s$^{-1}$.
There is no clear indication of a swirling motion of the bright points,
but one can imagine that magnetic flux concentrations trapped in a photospheric
swirl similar to the ones detected by \citet{ost_bonet+al2008,ost_bonet+al2010} 
may start to rotate and lead to chromospheric disturbances that would be similar
to the ones observed by \citet{ost_wedemeyer+rouppe2009} and similar as predicted by 
\citet{ost_nordlund1985}.

On a considerably larger scale but also in quiet Sun regions, \citet{ost_zhang+liu2011} 
observe extreme-ultraviolet cyclones in all EUV channels of the Atmospheric
Imaging Assembly (AIA) on board the Solar Dynamics 
Observatory (SDO). These cyclones seem to be rooted in rotating network magnetic 
fields that are simultaneously observed in the photosphere with the 
Helioseismic and Magnetic Imager (HMI; \citet{ost_schou+al2011}). They can
last for several to more than 10 hr and are found to be associated with EUV
brightenings (microflares) and EUV waves in their later phase. At this
stage it is unclear if and what kind of connections between 
\citeauthor{ost_wedemeyer+rouppe2009}'s swirl and \citeauthor{ost_zhang+liu2011}'s cyclones
may exist.

\subsection{Numerical Simulations}
\label{sec:numerical_simulations}
\citet{ost_shelyag+al2011} carried out magnetic and non-magnetic three-dimensional numerical 
simulations of solar granulation with the MURaM code \citep{ost_voegler2003,ost_voegler+al2005}
to analyze the generation of small-scale vortex motions 
in the solar photosphere. Starting with a magnetic field-free model, they observe the 
generation of swirls in the upper part of their model, as soon as a uniform, vertical 
magnetic field of 200~G strength is introduced. They find that the vortices in the upper 
photosphere are co-spatial with the magnetic field concentrations in the intergranular 
network in the lower photosphere that develop a short time after introduction of the 
magnetic field. Previously already, \citet{ost_voegler2004} found such a relation as well.
It is of course tempting to identify these vortical flows, which obviously occur in 
connection with magnetic flux concentrations, with the chromospheric swirls of
\citet{ost_wedemeyer+rouppe2009}. But for this, maps of \ion{Ca}{i} 854.2~nm
would need to be synthesized from the simulation for a more direct comparison.

\begin{figure}[]
\centerline{\includegraphics[width=0.70\textwidth]{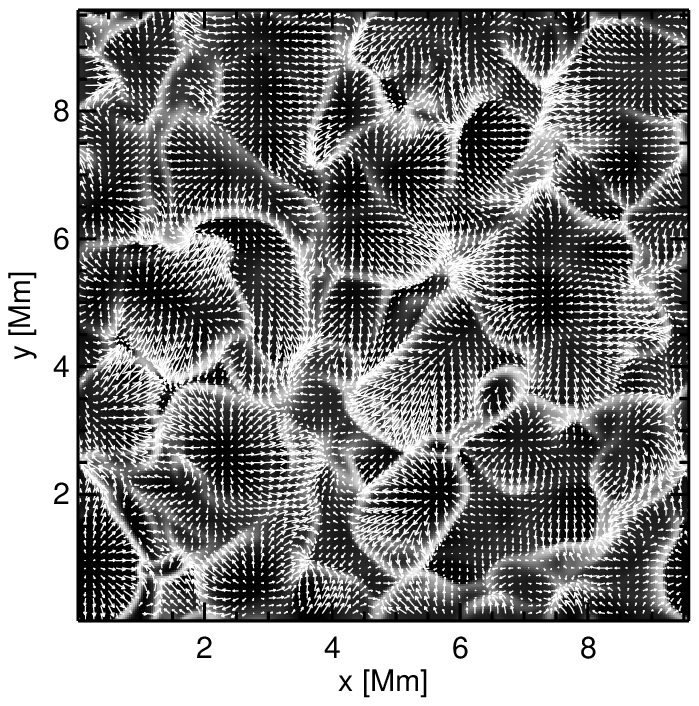}}
\vspace*{-1.7em}
\centerline{\includegraphics[width=0.70\textwidth]{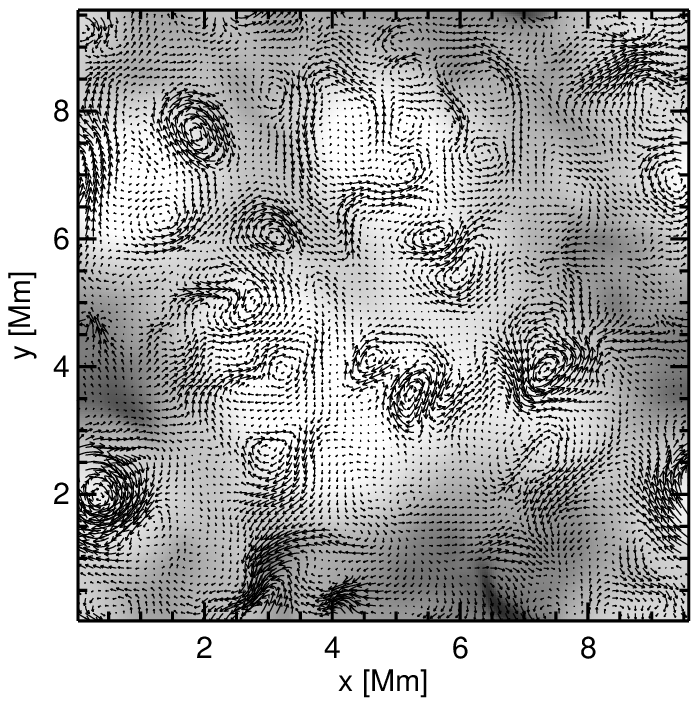}}
\caption{Time instances of the velocity field projected into the horizontal plane at 1300~km
above $\langle\tau_c\rangle = 1$. Top: Magnetic field-free simulation B0. Overplotted in
gray scale is the temperature. Temperature peaks delineate the meshwork of shock fronts. 
Bottom: Same time instance from the simulation, v50, which started with a homogeneous, vertical 
magnetic field of 50~G. Gray scales in the background show the logarithm of the field
strength, $\log |\mbox{\boldmath$B$}|$ from 1 to 100~G. The flow field is strikingly different 
from the field-free simulation and shock waves are absent.
Longest arrows with a 1\% area coverage have $v \ge 14$~km\,s$^{-1}$ (top) and
$v \ge 12$~km\,s$^{-1}$ (bottom).
\label{fig:swirl_co5bold}}
\end{figure}

As an independent, qualitative confirmation of the results of \citet{ost_shelyag+al2011}, we 
show in Figure~\ref{fig:swirl_co5bold} swirling motion that occurs in 
the upper part of a model atmosphere obtained from a simulation with the CO5BOLD 
code \citep{ost_freytag+al2012,ost_beeck+al2012}. We have carried out three equivalent simulation 
runs with a box size of $9.6 \times 9.6$~Mm$^2$ and a depth of 2.8~Mm, where the average 
depth of $\tau_c = 1$ is at mid height. Figure~\ref{fig:swirl_co5bold} (top) shows a time 
instance of the velocity field as projected into the horizontal plane at a height of 1300~km 
above $\langle\tau_c\rangle = 1$ for the simulation run B0 without magnetic fields. 
Added in gray-scale is the temperature, which shows a meshwork of sharp temperature peaks,
corresponding to shock fronts. The cellular flow field does not correspond to the
granular flow field in the deep photosphere---rather it is an independent, rapidly
evolving pattern at chromospheric heights \citep{ost_wedemeyer+al2004}. The longest arrows
correspond to a velocity of about 14~km\,s$^{-1}$. No swirls are visible in this 
magnetic field-free simulation. Figure~\ref{fig:swirl_co5bold} (bottom) shows the 
corresponding flow at the same time instance from an equivalent simulation, v50,
that started with a homogeneous vertical magnetic field of strength 50~G. The
flow field is conspicuously different and the meshwork of shock waves is absent. 
Swirling, or undulatory motion is now omnipresent and must have been introduced through 
the magnetic field. A third equivalent simulation, h50, was carried out in which 
horizontal field of 50~G strength was advected by updrafts through
the bottom boundary, similar to \citet{ost_steiner+al2008} or \citet{ost_stein+nordlund2006}.
This simulation shows again a flow pattern similar to the field-free case shown
in Figure~\ref{fig:swirl_co5bold} (top) because in this case the magnetic field is
more turbulent and complex and its average strength drops exponentially with height 
because of the many intricate loops that develop. These differences suggest that 
strong swirls in the upper atmosphere may best develop in case of a rather unipolar 
open field configuration as may occur at the base of a coronal hole.

In another independent study, \citet{ost_moll+al2012} carry out comparative simulations, 
similar to those shown in Fig.~\ref{fig:swirl_co5bold}. They find a considerably 
different flow structure in the upper photospheric layers of two simulations: 
the non-magnetic simulation is dominated by a pattern of moving shock fronts while 
the magnetic simulation shows vertically extended vortices associated with magnetic
flux concentrations. They note that both kinds of structures induce substantial 
local heating, in the magnetic case through Ohmic dissipation associated with the
swirling motion.

\citet{ost_shelyag+al2011} also derive the equation for the vorticity 
$\mbox{\boldmath$\omega$} = \nabla\times\mbox{\boldmath$v$}$
by taking the curl of the MHD momentum equation:
\begin{eqnarray}
\lefteqn{
\frac{{\rm D}\mbox{\boldmath$\omega$}}{{\rm D}t} = 
\overbrace{\rule{0pt}{15pt}(\mbox{\boldmath$\omega$}\cdot\nabla)\mbox{\boldmath$v$} 
- \mbox{\boldmath$\omega$}(\nabla\cdot\mbox{\boldmath$v$})
}^{\mbox{\footnotesize tilting and streching $\;T_1$}}
+\overbrace{\frac{1}{\rho^2}\nabla\rho\times\nabla p_{\rm gas}
}^{\mbox{\footnotesize baroclinic term $\;T_2$}} }\nonumber \\
&&+\overbrace{\frac{1}{\rho^2}\nabla\rho\times\left[\nabla p_{\rm mag} 
- \frac{1}{4\pi}(\mbox{\boldmath$B$}\cdot\nabla)\mbox{\boldmath$B$}\right]
}^{\mbox{\footnotesize magnetic baroclinic term $\;T_3$}}
+\overbrace{\frac{1}{4\pi\rho}\nabla\times(\mbox{\boldmath$B$}
\cdot\nabla)\mbox{\boldmath$B$}\rule{0pt}{17pt}
}^{\mbox{\footnotesize magnetic tension $\;T_4$}}\;,
\end{eqnarray}
and study the individual sources for vorticity, $T_1$ to $T_4$, in their 
simulation. Here, \hbox{\boldmath$v$} is the velocity, $\rho$ the density, 
$p_{\rm gas}$ the gas pressure, and \hbox{\boldmath$B$} the magnetic field.
They found that in the upper photosphere, the magnetic tension term
is most important, while beneath the surface of optical depth unity,
$\tau_c = 1$, the baroclinic term $|T_2|$ dominates $|T_4|$. Near 
$\tau_c = 1$, $|T_2|$ and $|T_4|$ are of similar magnitude. The other
terms are of less importance.

\begin{figure}[]
\centerline{\includegraphics[width=1.0\textwidth]{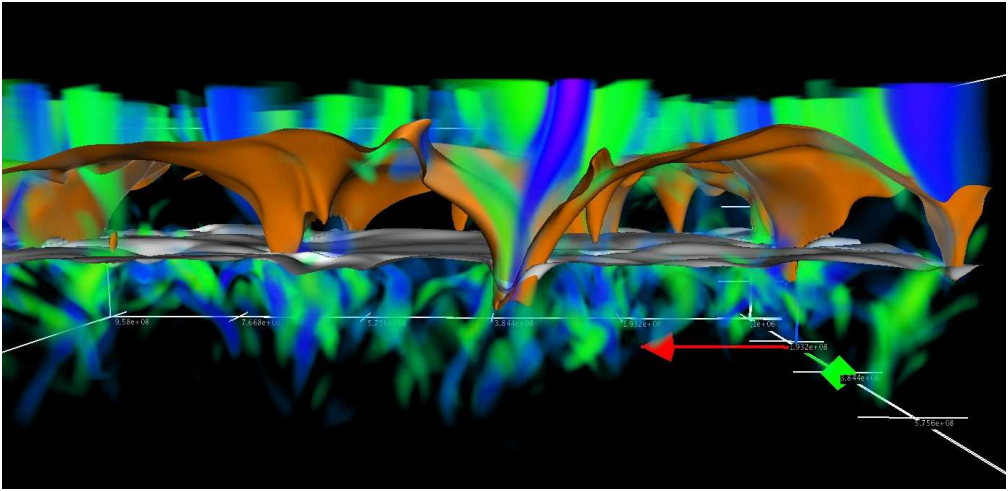}}
\caption{Vertical component of the vorticity, 
$\mbox{\boldmath$\omega$}_z = (\nabla\times\mbox{\boldmath$v$})_z$ 
(green and blue for opposite signature) together with the surface (brown) of 
plasma $\beta = 1$. At the locations of magnetic flux concentrations, the  
$\beta = 1$ surface reaches through funnels into the deep photosphere and even 
below the surface (gray) of optical depth unity. In 
between the funnels, it forms a horizontally extending canopy at chromospheric
levels. Above the $\tau_c = 1$ surface, vorticity is mainly confined
to the region of $\beta << 1$. Below it, in the convection zone, vorticity 
is virtually everywhere present below intergranular lanes. The simulation
was carried out with the CO5BOLD code \citep{ost_freytag+al2012}, the visualization
with VAPOR 3D \citep{ost_clyne+al2007}. Courtesy, Ch.~Nutto.
\label{fig:swirl_co5bold3d}}
\end{figure}

Since the plasma $\beta$ (ratio of thermal to magnetic pressure) generally 
decreases with height and assumes values smaller than one in the 
chromosphere and within magnetic flux concentrations in the photosphere, 
the results of \citet{ost_shelyag+al2011} indicate that vorticity may
be generated in regions of small $\beta$ through the magnetic tension term, 
i.e.\ by Lorentz forces. In fact, Figure~\ref{fig:swirl_co5bold3d} supports
this conjecture. At the location of magnetic flux concentrations,
the  $\beta = 1$ surface (brown) reaches through funnels deep into the 
photosphere and even below the surface (gray) of optical depth unity. Above
the surface of optical depth unity, vorticity (blue and green) is confined to 
within these funnels where $\beta << 1$. There is little vorticity outside
the funnels in the photosphere, where $\beta >> 1$. Clearly, this enhanced 
vorticity within the funnels must be generated by the magnetic field, possibly 
through the vortical motion flux concentrations are forced to follow when
getting trapped within intergranular swirling sinkholes of the type 
found by \citeauthor{ost_bonet+al2008}\ or \citeauthor{ost_wang+al1995}

\citet{ost_kitiashvili+al2011} discover in their (field-free) simulations tiny circular 
`density holes' in horizontal cross sections close to the solar surface.
These `density holes' with a density deficit of up to 60\% are typically located 
at vertices of intergranular lanes but sometimes they also occur in the middle 
stretch of an intergraular lane. They are associated with a deficit in temperature 
of about 20\%, a downdraft of up to 7~km\,s$^{-1}$, and swirling motion with
often supersonic horizontal velocities. \citeauthor{ost_kitiashvili+al2011}'s 
``whirlpools'' must be the same object that \citet{ost_nordlund1985} found to be
a consequence of the centripetal force associated with a vortical flow.
Underdense near-surface vertical vortices of similar properties are also 
found by \citet{ost_moll+al2011} from magneto-convection simulations with the MURaM 
code \citep{ost_voegler2003,ost_voegler+al2005}. According to these authors, they cause 
a local depression of the optical surface, which would potentially be observable 
as bright points in the dark intergranular lanes. 

Most important,
\citet{ost_kitiashvili+al2011} also found that whirlpools can attract and capture other 
swirls of opposite vorticity. They showed that this processes of vortex interaction, 
can cause the excitation of acoustic waves on the Sun. They conclude that strongly 
interacting vortices in the top layers of the convection zone, in particular the 
interaction of the vortices with opposite sign of vorticity, may play
an important role in the excitation of solar acoustic oscillations. In their
simulations, they did not find a preference in directions of the vortex rotation,
which should be expected since neither the Coriolis force nor differential rotation
is included. However, if the findings of a preferred sense of rotation of 
\citet{ost_bonet+al2010} should substantiate and if theses swirls are indeed an important
source of acoustic oscillations, one might speculate that this could have
observable consequences for the p-mode oscillations. One could possibly expect 
reduced power at latitudes of enhanced differential rotation.

\begin{figure}[]
\centerline{\includegraphics[width=0.7\textwidth]{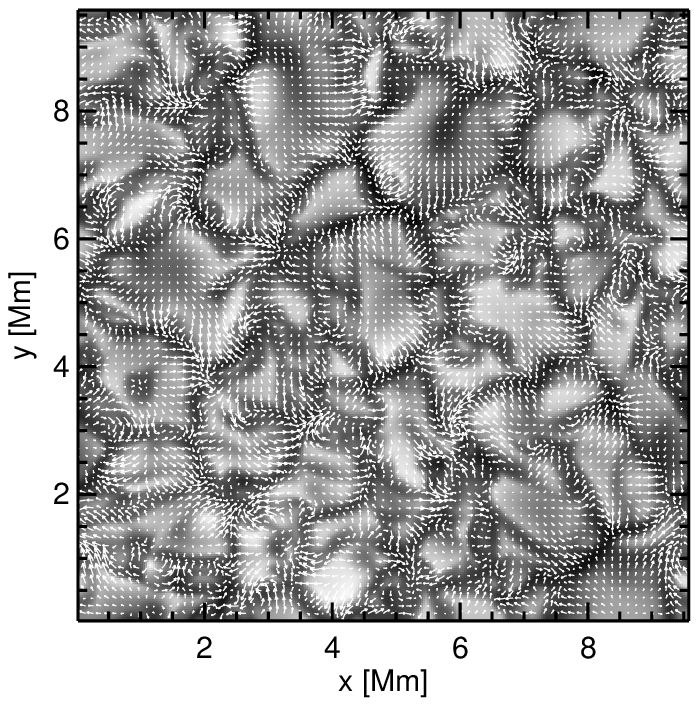}}
\vspace*{-1.7em}
\centerline{\includegraphics[width=0.7\textwidth]{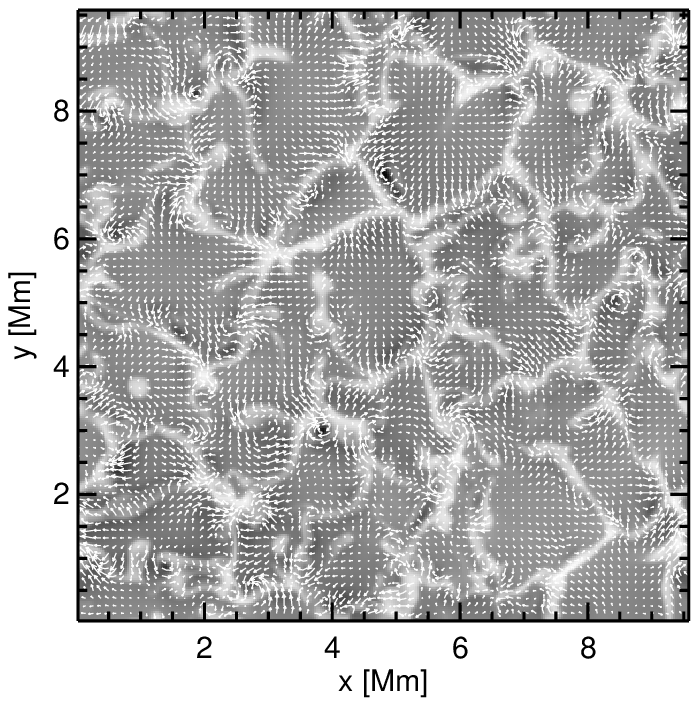}}
\caption{Top: Time instance of the bolometric intensity and the velocity field 
projected into the horizontal plane located at $\langle\tau_c\rangle = 1$. 
Bottom: Corresponding time instance of the logarithm of the density 
($-6.25 \le \log\rho \le -6.86$) and the velocity field 
projected into the horizontal plane located 120~km below $\langle\tau_c\rangle = 1$.
Underdense vortices ar visible at, e.g., $(x,y) = (3800,3200)$~km, 
$(x,y) = (5000,8500)$~km, $(x,y) = (1900,8250)$~km, and
twin swirls at $(x,y) = (5000,7000)$~km. They have a density of typically
60\% of the mean density. Intergranular lanes have typically 115\%. The snapshot
is from a magnetic field-free simulation with CO5BOLD.
\label{fig:surfaceflow_co5bold}}
\end{figure}

Figure~\ref{fig:surfaceflow_co5bold} (top) shows the bolometric intensity of a 
simulation snapshot of the above mentioned field-free CO5BOLD simulation B0.
Superimposed on the intensity map is the velocity field projected into 
the horizontal plane at the average optical depth unity. This is approximately,
what could be derived from observations with the help of correlation tracking
and equivalent techniques. The figure demonstrates, that swirling motions are 
not particularly conspicuous. In most vertices of inter-granular lanes, there 
is hardly any sign of swirling motion visible in this snapshot. 
Only at $(x,y) = (6000,3000)$~km, there is clear evidence of a `whirlpool'. 
The situation changes drastically when examining the density at about 100~km 
below mean optical depth unity, which section is shown in 
Figure~\ref{fig:surfaceflow_co5bold} (bottom). There, we can see
small underdense circular areas with a diameter of $\approx$ 100--200~km,
which are associated with swirling motion, e.g., at 
$(x,y) = (3800,3200)$~km, $(x,y) = (5000,8500)$~km, $(x,y) = (1900,8250)$~km,
and twin swirls at $(x,y) = (5000,7000)$~km. They have a density of typically
60\% of the mean density, while the intergranular lanes have typically 115\%.
They also are typically 25--35\% cooler than the average temperature.
These findings qualitatively corroborate corresponding results by 
\citet{ost_kitiashvili+al2011} and \citet{ost_moll+al2011}. 
We do not see a bright point in the bolometric
intensity map at the location of the whirlpool as was reported by 
\citet{ost_moll+al2011} but this may be due to insufficient spatial resolution
of the present simulation. There is no one-to-one correspondence of swirls
visible at the solar surface and underdense vortices 120~km below it. For
example, the conspicuous swirl visible at $(x,y) = (6000,3000)$~km at the 
surface has no underdense counterpart 120~km beneath.
On the other hand, the underdense swirl at $(x,y) = (3800,3200)$~km seems
to be associated with an intensity enhancement at the surface but not with
a clear vortical flow. The twin swirls at $(x,y) = (5000,7000)$~km are 
associated with a granular lane visible in the intensity map 
(Fig.~\ref{fig:surfaceflow_co5bold} top). In fact, it looks like the two 
`footpoints' of the horseshoe-like granular lane were leading into the twin 
sinkholes, forming all together a loop-like vortex tube. Granular lanes 
and vortex tubes are reviewed in the following subsection.

\subsection{Granular Lanes and Horizontal Vortex Tubes}
\label{sec:vortex_tubes}
\begin{figure}[!h]
\centerline{\includegraphics[width=0.8\textwidth]{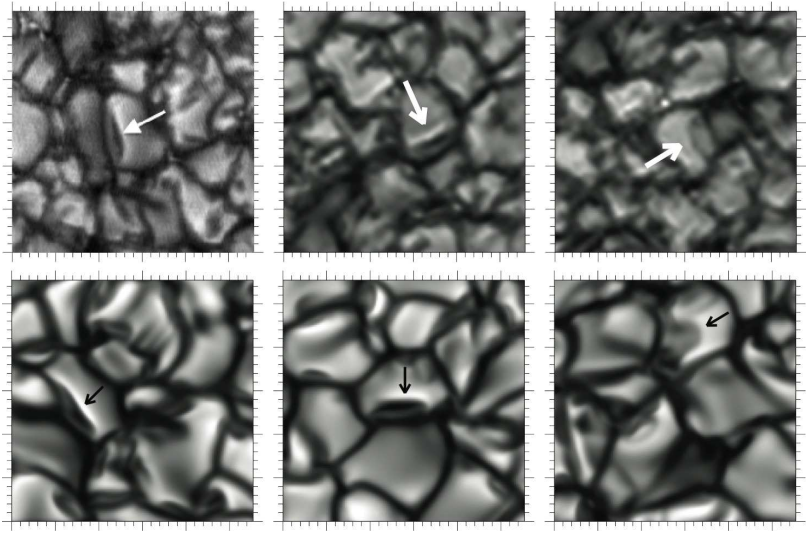}}
\caption{Top row: Granules of the visible solar surface in fields of view 
of approximately $5.6 \times 5.6$~Mm as observed with 
the SST (1st panel, courtesy L.~Rouppe van der Voort) and
the Sunrise ballon-borne telescope (2nd and 3rd panel). 
The threadlike, bright/dark stripes, marked with arrows, develop at the 
edge of a granule and move into the granule itself. 
Bottom row: Equivalent sections from an MHD simulation with the CO5BOLD code.
The arrows mark strikingly similar threadlike objects, which prove to be
vortex tubes. Adapted from \citet{ost_steiner+al2010}.
\label{fig:vortex_tubes_fig1}}
\end{figure}

It was known for long that in the visible continuum, the 
edges of granules tend to be brighter than the interior of granules or the
intergranular  space. High resolution observations have refined this
knowledge. It was found that granules frequently show substructure in the form
of lanes composed of a leading bright rim and a trailing dark edge, which form
at the boundary of a granule and move together into the granule itself.
Fig.~\ref{fig:vortex_tubes_fig1}
shows in the top row granules of the visible solar surface in fields of 
view of 5600 x 5600 km as observed with the SST and the Sunrise ballon-borne 
telescope. 
Events of granular lanes are marked with arrows. Simultaneously recorded
Doppler maps \citep[see][]{ost_steiner+al2010} show that material flows in the 
upward direction within the bright granular lane with speeds of up to
1.8~km\,s$^{-1}$. Most of the area behind the bright rim, including the 
dark lane, harbors upflows as well but less strong, and sometimes even
downflows. Close to the border of the granule and within the adjacent 
intergranular lane, the plasma flows in the downward direction with speeds
up to 1.2~km\,s$^{-1}$.

The bottom row of Fig.~\ref{fig:vortex_tubes_fig1} shows equivalent sections 
from a computer simulation. The arrows mark strikingly similar threadlike 
objects as are observed. From cross sections through the computational domain 
of the simulation,  \citet{ost_steiner+al2010} conclude that these granular lanes 
are the visible signature of horizontally oriented vortex tubes. The
axis of the vortex tube typically coincides with the trailing dark edge.
Directly above the vortex tube, the flow assumes transonic speeds, roughly 
parallel to the solar surface.
There, gas pressure and temperature are low, which opens a relatively
transparent view to the cool interior of the vortex tube causing the dark 
edge. Rather than a proper movement, it is the shape and size of the vortex 
tube which changes in time and causes the horizontal displacement of the 
bright lane and the expansion of the trailing darkish area. 
In the simulation, we see occasional magnetic field intensification
at locations of vortex tubes---one such case is shown in
\citet{ost_steiner+al2011}. Whenever magnetic fields are present, they tend to
point in the horizontal direction near or above the $\tau_c = 1$ surface.

\citet{ost_yurchyshyn+al2011} observe that intergranular jets, originating in the 
intergranular space surrounding individual granules \citep{ost_goode+al2010}, tend 
to be associated with the formation and evolution of granular lanes.
They speculate that the intergranular jets may result from the interaction 
of the turbulent small-scale fields associated with the vortex tube with
preexisting larger-scale fields in the intergranular lanes. Since
we found in Fig.~\ref{fig:surfaceflow_co5bold} a singular case where 
twin whirlpools beneath the surface were an integral part of a granular lane,
it cannot be excluded that sinkholes are an additional important ingredient for
the production of the intergranular jets.

\subsection{Small-Scale Vortices in Simulations of Solar Surface Convection}
\label{sec:swirling_strength}
A synoptical view of small-scale vortices in simulations of solar surface 
convection was provided by \citet{ost_moll+al2011}. These authors define the
``swirling strength'' based on an eigenanalysis of the velocity gradient 
tensor. It allows them to pick high-vorticity regions in which the plasma 
is swirling. The swirling regions form an unsteady network
of highly tangled filaments, some of which protrude above the
optical surface.

\begin{figure}[t]
\centerline{\includegraphics[width=1.0\textwidth]{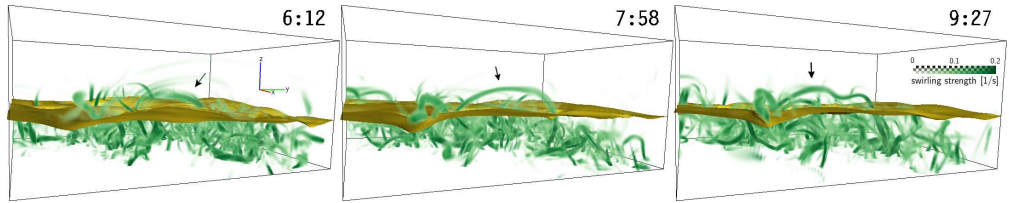}}
\caption{Rise and fall of a vortex arc. The plots display the swirling 
strength (green volume rendering) and the optical surface (yellow) at
three different times (labels are in minutes). The size of the box 
shown is $1.5 \times 1.5 \times 0.8$~Mm$^3$. Credit: \citet{ost_moll+al2011}
reproduced with permission \copyright\ ESO.
\label{fig:swirl_moll}}
\end{figure}

Near the optical surface, vertically oriented swirls are preferentially
located within the intergranular lanes, where cooled fluid is sinking 
down in a turbulent fashion. Horizontal swirls are predominant at the 
edges of the granules. Above the optical surface, the three-dimensional
structure of swirls is manifold, but often they appear in the form of 
bent and arc-shaped filaments as shown in Fig.~\ref{fig:swirl_moll}.
A similar type of structure was also reported by \citet{ost_stein+nordlund1998}
and by \citet{ost_muthsam+al2010}. \citet{ost_kitiashvili+al2012} discuss
vortex tubes that reach into the chromosphere
and interact with magnetic fields in their 3D simulations and they conclude
that magnetized vortex tubes would effectively transfer 
energy and momentum 
from the photosphere to the chromosphere.

From Fig.~\ref{fig:swirl_moll}, one can estimate that the arc-shaped
swirl has a diameter of about 30~km, which is close to the grid size
of 10~km of the numerical simulation. One could therefore speculate 
that in reality the swirling
filaments would be even thinner. It would be extremely difficult to
detect them with present day solar telescopes. Doppler-shift measurements
have a depth resolution, limited by the formation width of
the spectral line or a combination of spectral lines, which is 
definitively larger than 30~km.

\section{The Horizontal Magnetic Field of Quiet Sun Regions}
A fundamental achievement attained with Hinode's Solar Optical 
Telescope/Spectro-Polarimeter (SOT/SP) \citep{ost_sakurai2008} was the discovery 
that in the mean, the magnetic field 
over wide areas of the quiet Sun photosphere is predominantly horizontal  
\citep{ost_lites+al2007,ost_lites+al2008,ost_orozco-suarez+al2007,ost_danilovic+al2010}.
From the analysis of time-averaged deep mode Stokes spectra
with an effective integration time of 67.2~s and a noise level in the 
polarization continuum of about $2.9\times 10^{-4}\,I_c$,
\citet{ost_lites+al2008} came to the conclusion that the spatially
averaged horizontal apparent flux density, as derived from 
wavelength-integrated measurements of the Zeeman-induced linear polarization,
is 55\,Mx\,cm$^{-2}$, five times more than the corresponding average vertical 
apparent flux density of 11\,Mx\,cm$^{-2}$. This result, however,
remained not undisputed.

\subsection{Critiques of the Horizontal Field Measurements}
\label{sec:critiques}
\citet{ost_martinez-gonzalez+al2008} analyzed spectro-polarimetric data of the
\ion{Fe}{i} lines at 1.5648\,$\mu$m and 1.5652\,$\mu$m, recorded with the
Tenerife Infrared Polarimeter (TIP) at the
Vacuum Tower Telescope (VTT, Observatorio del Teide), of very quiet Sun
regions at various heliocentric distances on the solar disk from 
$\mu = 1.0$ to $\mu = 0.28$. They restrict the analysis
to profiles with a degree of polarization, $\sqrt{Q^2 + U^2 + V^2}$ larger
than $4 \times 10^{-4}\, I_c$. They found that the circular 
and linear polarization amplitudes do not have any clear dependence on 
the heliocentric angle. This result points to an isotropic distribution
of magnetic fields and is against any field topology with a preferred
orientation within the field-of-view.

\citet{ost_beck+rezaei2009} observed a quiet Sun region at disk center with
TIP{\at}VTT, using the same lines as \citet{ost_martinez-gonzalez+al2008}.
The noise level of the polarization signal was $2\times 10^{-4}\,I_c$, where
they set a threshold of $1\times 10^{-3}\, I_c$ for the polarization degree
during the analysis. They find that the total magnetic flux contained in the 
more inclined to horizontal fields ($\gamma > 45\deg$) is about two times
smaller than that of the less inclined fields. However, they
do not conclude that this result would necessarily contradict the results of
\citeauthor{ost_lites+al2008}\ because of the different formation heights of
the lines at 1.56\,$\mu$m and at 630.2\,nm. In fact, simulations
\citep{ost_schuessler+voegler2008,ost_steiner+al2008,ost_danilovic+al2010} 
show a strong height
dependence of the horizontal magnetic field---while the vertical component
may dominate in the deep photosphere, the horizontal component becomes
important in the upper photosphere and lower chromosphere only.
\citet{ost_beck+rezaei2009} also emphasize that the pixel to pixel variation
of the thermal structure of the atmosphere should be properly taken
into account during the analysis of the polarization signal. This is
not taken into account when using a global (pixel independent) calibration 
curve for the derivation of magnetic field strengths as was done by
\citeauthor{ost_lites+al2008}

\citet{ost_asensio-ramos2009} carried out a Bayesian analysis of the `normal mode'
data (4.8~s exposure time, $\sigma_{\rm noise}\approx 1.2\times 10^{-3}I_c$) 
of \citet{ost_lites+al2008}. He points out
that the noise level present in the Hinode SOT/SP observations induces a 
substantial loss of information for constraining the angular distribution
of the magnetic field. However, the results indicate that the field of 
pixels with small polarimetric signals has a 
quasi-isotropic distribution. He also finds that the magnetic
field strength in the internetwork region is clearly in the hectogauss 
regime with 95\% confidence. 

Yet another way of analyzing the Hinode SOT/SP data was advanced by
\citet{ost_stenflo2010}. He carried out a thorough analysis of the ratio
of the amplitudes of the blue wing of Stokes $V$ of \ion{Fe}{i} 
$\lambda\lambda$ 630.15 and 630.25\,nm and finds a ``magnetic dichotomy'' 
with two distinct populations, representing strong (kG) and weak fields.
With regard to the inclination he finds that the angular distribution is 
extremely peaked around the vertical direction for the largest flux 
densities, but gradually broadens for smaller flux densities, to become 
asymptotically isotropic at zero flux density. This means that the
magnetic field has a predominantly vertical orientation, quite opposite
to the findings of \citet{ost_lites+al2007,ost_lites+al2008} and 
\citet{ost_orozco-suarez+al2007}. We come back to Stenflo's method in 
Sect.~\ref{sec:line_ratio}.

\begin{table}[t]

\caption{Angular distribution of the internetwork magnetic field as
determined by various authors. The entries in the column `angular 
distribution' must be considered an over-all summary, the details 
being more complex. A pure isotropic angular distribution would lead 
to  $\langle B^{\mathrm{T}}_{\mathrm{app}}\rangle/
\langle B^{\mathrm{L}}_{\mathrm{app}}\rangle = \pi/2$. The entries 
in the lower part of the table refer to numerical simulations, for
which the ratio $\langle B^{\mathrm{T}}_{\mathrm{app}}\rangle/
\langle B^{\mathrm{L}}_{\mathrm{app}}\rangle$ was derived from
synthesized Stokes profiles and application of a PSF corresponding
to that of Hinode/SOT. Numbers in parentheses without PSF. 
\label{tab:angular_distribution}}
\smallskip
\begin{center}
{\small\footnotesize
\begin{tabular}{rllrll}
\tableline
\noalign{\smallskip}
no.& authors & instrument/& line &      angular &
$\langle B^{\mathrm{T}}_{\mathrm{app}}\rangle/$\\[0ex]
   &         & simulation &  [nm]& distribution & $\langle |B^{\mathrm{L}}_{\mathrm{app}|}\rangle$\\
\noalign{\smallskip}
\tableline
\noalign{\smallskip}
 1&\citet{ost_lites+al2007,ost_lites+al2008}  & SOT/SP & 630& predominantly horizontal & 5    \\
 2&\citet{ost_orozco-suarez+al2007}           & SOT/SP & 630& predominantly horizontal & 2.1  \\
 3&\citet{ost_martinez-gonzalez+al2008}       & VTT/TIP&1560& isotropic distribution & --- \\
 4&\citet{ost_beck+rezaei2009}                & VTT/TIP&1560& predominantly vertical& 0.42\\
 5&\citet{ost_asensio-ramos2009}              & SOT/SP & 630& isotropic for weak fields & ---\\
 6&\citet{ost_danilovic+al2010}               & SOT/SP & 630& predominantly horizontal & 5.8\\
 7&\citet{ost_stenflo2010}                    & SOT/SP & 630& predominantly vertical   & ---\\
 8&\citet{ost_ishikawa+tsuneta2011}           & SOT/SP & 630& predominantly vertical& 0.86\\
 9&\citet{ost_borrero+kobel2011}              & SOT/SP & 630& undeterminable& ---\\
10&\citet{ost_borrero+kobel2012}              & SOT/SP & 630& non-isotropic& ---\\
\noalign{\smallskip}
\tableline
\noalign{\smallskip}
11&\citet{ost_steiner+al2008}                 & h20    & 630& predominantly hor- & 4.3 (2.8) \\
  &                                           & v10    & 630& izontal            & 1.6 (1.5) \\
12&\citet{ost_danilovic+al2010}               & C mf=3 & 630& predominantly hor- & 9.8 (3.5) \\
  &                                           & C+$B_{\rm ver}$ & 630& izontal   & 4.2 (2.6) \\
\noalign{\smallskip}
\tableline
\end{tabular}
}
\end{center}
\end{table}

\citet{ost_ishikawa+tsuneta2011} find, based on Hinode SOT/SP data,
a clear positional association between the vertical and the horizontal 
magnetic fields in the internetwork region. Essentially, all of the 
horizontal magnetic patches are associated with vertical magnetic patches 
and half of the vertical magnetic patches are associated with horizontal 
magnetic patches. This points to small-scale magnetic loops with bipolar 
footpoints, and to canopy fields of magnetic flux concentrations as an 
important source of the horizontal magnetic field. In fact, tiny emerging
loops of granular and sub-granular scale were observed with SOT/SP by 
\citet{ost_ishikawa+al2010} and \citet{ost_martinez-gonzalez+al2010}. 
\citet{ost_steiner+al2008} show an example of an emerging sub-granular 
horizontal field patch framed by opposite polarity footpoints of
vertical field from their simulations. They describe it as a consequence
of convective overshooting, which transports horizontal fields into
the photosphere. \citet{ost_schuessler+voegler2008} see the loopy structure
of the small scale magnetic field in the photosphere as a direct outcome 
of the surface dynamo which operates in the surface layers of the 
convection zone (see also \citet{ost_steiner2010} for a sketch explaining
why such a loopy structure leads to an increasing dominance of the horizontal 
field component with increasing height in the atmosphere).  Prior to this,
\citet{ost_lites+al1996} report isolated, predominantly horizontal magnetic flux 
patches of $1\arcsec$--$2\arcsec$ and smaller size, often occurring between 
regions of weak opposite polarity Stokes $V$ profiles, which they suggested
to be the emergence of small, concentrated loops of magnetic flux.

\citet{ost_ishikawa+tsuneta2011} measure for an internetwork region a mean
longitudinal (vertical) magnetic flux density of 8.3~Mx\,cm$^{-2}$
and a mean transverse (horizontal) flux density of 7.1~Mx\,cm$^{-2}$,
thus a clear dominance of the vertical fields.
Note that while the former number is close to the value obtained
by \citet{ost_lites+al2007,ost_lites+al2008} and \citet{ost_orozco-suarez+al2007}, the
latter is about 8 times smaller. One reason for this discrepancy is the
stringent noise criteria that was applied to the data by 
\citet{ost_ishikawa+tsuneta2011}. Since a transversal field produces
a much weaker signal in linear polarization than
a longitudinal field of equal strength in circular polarization,
there is a bias towards longitudinal fields when applying a
common noise level and discarding any signal below it. Correspondingly, 
\citet{ost_ishikawa+tsuneta2011}
find only 10.1\% of a supergranular cell sized subfield
having sufficient linear polarization for entering the analysis, while
43.8\% have sufficient circular polarization. On the other hand 
applying no noise level gives preference to transversal fields 
because pure noise would produce spurious large amounts
of transversal field. We come back to the problem of
selection effects in Sect.~\ref{sec:selection}.

Table~\ref{tab:angular_distribution} provides an overview of the above discussed  
various attempts to determine the horizontal vs.\ the vertical mean flux density (or the
angular distribution) of the internetwork magnetic field.

\subsection{The Photon-Noise Problem}
\label{sec:photon_noise}
\begin{figure}[t]
\centerline{\includegraphics[width=0.75\textwidth]{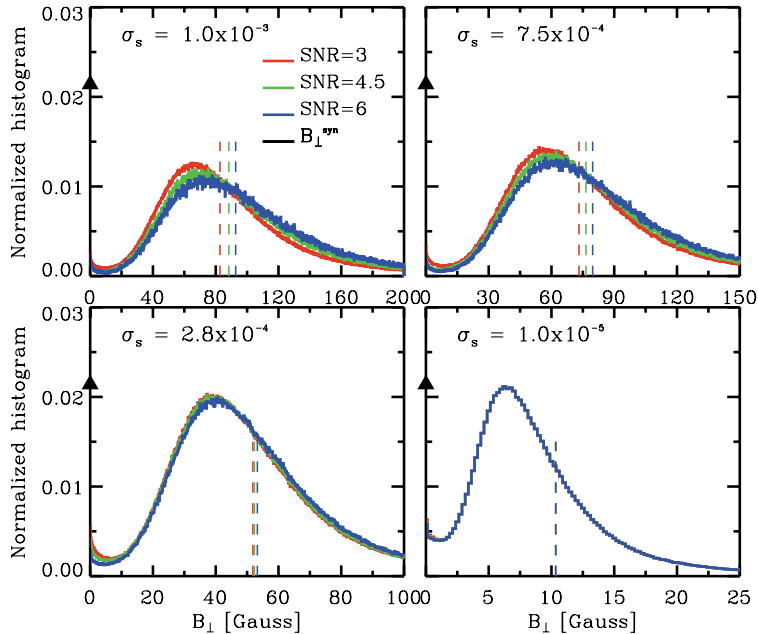}}
\caption{Histograms of the transverse component of the magnetic
field, $B_{\perp}$, which is purely induced by different levels of 
noise, $\sigma_s$, indicated in the top left of each panel. The
different curves correspond to different signal-to-noise threshold 
for a  pixel to be considered for inversion.
The vertical black arrow indicates the $\delta$-Dirac distribution
of the true, original transverse component. The vertical, dashed
lines indicate the center of gravity of the histograms. Credit:
\citet{ost_borrero+kobel2011} reproduced with permission \copyright\ ESO.
\label{fig:borrero+kobel_fig9}}
\end{figure}

The main reason for the disparate results highlighted in 
Table~\ref{tab:angular_distribution} is due to the different 
sensitivity of linear and circular polarization to magnetic fields, in 
combination with the finite sensitivity of the measurements due to photon 
noise. The problem was nicely demonstrated and discussed by 
\citet{ost_borrero+kobel2011}. In order to test if it was actually possible 
to accurately retrieve the magnetic field vector employing the \ion{Fe}{i} 
line pair at 630~nm in regions with very low polarization signals such as 
in internetwork regions, \citet{ost_borrero+kobel2011} carried out several 
Monte-Carlo simulations with synthetic data. They start from a map
of magnetic field vectors and associated atmospheric parameters, 
which they obtained in the first place from
a Stokes inversion of real SOT/SP data. Then, they arbitrarily set the 
transversal magnetic field to zero, synthesize the Stokes profiles from
this map and add different amounts of noise. Then they invert these data 
back to magnetic field vectors as if they were actually observational data. 
Clearly, the subsequent inversion should result in zero transversal field. However, 
this is not the case as is shown in Fig.~\ref{fig:borrero+kobel_fig9}.

The four panels show the results for different amounts of noise that 
was added to the synthesized profiles---from $1.0\times 10^{-3}\,I_c$ 
(top left) to $1.0\times 10^{-5}\,I_c$ (bottom right). 
$2.8\times 10^{-4}\,I_c$
corresponds to the noise level of the least noisy SOT/SP data used by 
\citet{ost_lites+al2007,ost_lites+al2008}. The inversion procedure asks for
a threshold of the signal to noise ratio. When either Stokes $Q$, $U$, or
$V$ are above this noise level, the pixel is considered
for inversion. This ratio was taken 3, 4.5, and 6,
but the different curves corresponding to these values are 
almost identical. The curves are histograms of the transversal
magnetic flux density and should really be $\delta$-Dirac 
distributions centered at zero if the inversion was correct. 
However, the center of gravity of the histograms (indicated
by the vertical dashed lines) is 80--90~G for a noise level
of $1.0\times 10^{-3}\,I_c$ and close to 55~G for the noise level
of $2.8\times 10^{-4}\,I_c$, which is, coincidence or not,
the value for the mean transversal apparent flux density
determined by  \citet{ost_lites+al2007,ost_lites+al2008}. Even at
a hypothetical noise level of $1.0\times 10^{-5}\,I_c$, the
retrieved histogram of the transversal field still has
a center of gravity of 10~G and peaks at about 6~G instead
of 0~G. 

Here is a note in place. The signal to noise limits
used here refer to single wavelength points, not to integral
quantities like 
$Q_{\rm tot} = \int_{\lambda_{\rm b}}^{\lambda_{\rm r}}
Q(\lambda)\,{\rm d}\lambda\, \Big/ \,
I_{\rm c}\!\int_{\lambda_{\rm b}}^{\lambda_{\rm r}}$.
Thus, even though the probability that photon noise produces a
signal above $3\sigma$, at a given wavelength position, is only 
0.3\%, it is almost guaranteed that one of the 112 wavelength
points in either Stokes $Q$, $U$, or $V$ surpasses this limit and
therefore the full profiles enter the analysis. Thus, even with 
a signal to noise ratio of 5, there are still ample profiles 
that enter the analysis with linear polarization that is 
interpreted as real although it would be strictly zero without 
the addition of noise. Also, most of these profiles are chosen
because at least one of the wavelength points in Stokes-$V$
surpasses the signal-to-noise limit and none in Stokes $Q$ or $U$.
This then
leads for example in the case of $\sigma = 2.8\times 10^{-4}\,I_c$
to a residual horizontal field strength of 55~G, much more
than the corresponding vertical field strength, which in 
this case is around 20~G and therefore, the field would
be interpreted as predominantly horizontal although it
is strictly vertical in reality. This devastating result
sheds serious doubts on previous determinations of the 
angular distribution of internetwork magnetic fields
based on Stokes inversion methods.

In defense of the work of \citet{ost_lites+al2007,ost_lites+al2008}
it should be noted however that they worked exclusively with
integral quantities, $V_{\rm tot}$ and $L_{\rm tot}$, which
should be much less affected by noise. However this forced them 
to use a previously determined calibration curve for the
conversion of total linear and total circular polarization 
into field strength. This calibration was derived from
a given atmosphere and therefore it cannot take pixel to
pixel variation of the atmospheric parameters into account,
which was found to be important by \citet{ost_beck+rezaei2009}.
Also, \citeauthor{ost_lites+al2008}\ did not use Stokes $Q$ and
$U$ separately, but tried to determine the 
``preferred-frame azimuth'' for each pixel so that
Stokes $U$ vanishes, enhancing the signal in Stokes $Q$.
This procedure should enhance the signal to noise in
linear polarization but is of course itself subject to noise. 
\citet{ost_danilovic+al2010} applied  \citeauthor{ost_lites+al2008}'s
method to white noise with a level of $\sigma = 8\times 10^{-4}\,I_c$
and obtained a mean spurious transversal magnetic field strength
of 36~G, which is about half of what is shown for a similar noise level 
in Fig.~\ref{fig:borrero+kobel_fig9}, indicating that 
\citeauthor{ost_lites+al2008}'s method produces considerably less
spurious horizontal field than straightforward Stokes
inversion does.

\subsection{Selection Effects}
\label{sec:selection}
There are different ways of selecting pixels as having good
enough signal for a trustworthy determination of the magnetic
field vector, all leading to biases in one or the other way,
again due to the largely different sensitivity of the circular
and linear polarization to magnetic fields.

Selecting pixels only which have either Stokes $V$ or $Q$ or
$U$ above the noise level gives a strong bias towards
horizontal fields. This is because most pixels will be selected
because they have a Stokes $V$ above the noise level (because
of the high sensitivity of Stokes $V$ to longitudinal fields)
but these pixels are dominated by noise in Stokes $Q$ and $U$,
which then will yield substantial amounts of false transversal 
field (because of the low sensitivity of linear polarization to 
transversal fields).

One could think that the obvious remedy consists in selecting
pixels only that have signals in Stokes $V$ and $Q$ or $U$
($V \wedge (Q\vee U)$) above the noise level. However, this
again gives preference to horizontal fields because in this
case only pixels with strong transversal fields are chosen,
while the many pixels with Stokes $V$ above the noise level
and $Q$ or $U$ below it are discarded.

One way (which has not been tried out so far) to circumvent these 
problems  would consist in setting limits in the real 
physical space of field strengths, instead of in the 
Stokes space. Thus, we demand that either $B_{\|}$ or
$B_{\perp}$ be larger or equal to a given field-strength
limit $B_{\rm lim}$. Then, 
\begin{equation}
  V_{\rm lim} = c_c B_{\rm lim}\quad\mathrm{and}\quad
  Q_{\rm lim} = c_l^2 B_{\rm lim}^2\;,
\end{equation}
where we now work without loss of generality in the
preferred-frame azimuth where $U$ vanishes. $c_c$
and $c_l$ are calibration constants that could possibly
vary form pixel to pixel. Then
\begin{equation}
  \frac{V_{\rm lim}}{c_c} = B_{\rm lim} = \sqrt{\frac{Q_{\rm lim}}{c_l^2}}
  = \frac{1}{c_l}\sqrt{n\sigma_{\rm noise}}\;,
\end{equation}
where we now demand that Stokes $Q$ be $n\sigma_{\rm noise}$
above the noise level if it is to be used for determining the
transversal field. This then leads to the selection criterium
\begin{equation}
\label{eq:criterium}
  Q \ge n \sigma_{\rm noise} \quad \mathrm{or} \quad
  V \ge \displaystyle\frac{c_c}{c_l}\sqrt{n\sigma_{\rm noise}}\;.
\end{equation}
Note that the criterium for Stokes $V$ is much more stringent than
that for $Q$, which is expression of setting equal lower field-strength 
limits to the transversal and longitudinal field. Still, we expect
many pixels to satisfy the stringent criterium for Stokes $V$ but still
not the less stringent criterium for Stokes $Q$. In this case we propose 
to discard $Q$, assuming the field to be strictly vertical.
On the other hand in the rare case where $Q$ satisfy the criterium
but not $V$, we consider the field as strictly horizontal. In
case where both $Q$ and $V$ satisfy their individual criterium,
the magnetic field vector can be determined.

These criteria should remedy the one-sided preference for
horizontal fields but still has its own problems. It can
be expected to give some preference to strictly vertical and
strictly horizontal fields. Also, large amounts of pixels that 
would have a $V$ signal above $n\sigma_{\rm noise}$ are 
discarded for inversion because they do not fulfill the more stringent
criterium for $V$, Eq.~\ref{eq:criterium}. Therefore, for only
a fraction of a given polarization map the magnetic field
could be retrieved, making it difficult to 
reliably determine  mean values of transversal and longitudinal
flux over the full map.

\subsection{Stokes-{\boldmath $V$} Amplitude Ratios}
\label{sec:line_ratio}
There are, roughly speaking, two ways of retrieving the magnetic
field vector from polarimetric data. With the \emph{Stokes inversion}
procedure one starts with a given atmospheric model including a 
magnetic field, computes the Stokes profiles of one or several
spectral lines by solving the Unno-Rachkowsky equations of polarized
radiative transfer in a forward fashion, and compares them to the
observed ones. Then, the magnetic field vector, thermal atmospheric 
structure, line-of-sight velocity, magnetic filling factor, and possibly
the micro and macro turbulence parameters and stray-light contributions 
are iteratively adjusted until a satisfactory agreement between
synthesized and observed profiles is achieved. The first such automatic
Stokes inversion code was described in \citet{ost_keller+al1990}. This 
procedure is repeated pixel by pixel of a polarimetric map, so that the 
atmospheric parameters are allowed to vary, say from the interior of a 
granule to the intergranular lane. Stokes inversion procedures were chosen 
by the authors of the second, fourth, fifth, and ninth row of 
Table~\ref{tab:angular_distribution}. In a sense, it can be considered
the classical way of atmospheric modeling as is performed in 
deriving standard stellar model atmospheres. 

Many solar physicist
distrust Stokes inversion procedures because of the many free
parameters that need to be adjusted (which renders the solution
non-unique), the potential vulnerability to noise, and the 
relatively complex programs, which give the impression of black boxes
(see \citet{ost_ruiz-cobo2007} for replies to such concerns).
Instead, they take resort in simple procedures derived from first 
principles of spectro-polarimetric line formation theory. This
approach may allow them to work with spectrally integrated quantities, 
like $V_{\rm tot}$ and $L_{\rm tot}$ mentioned in 
Sect.~\ref{sec:photon_noise}, which may be less susceptible to noise.
However, such basic receipts invariably 
end up in using a calibration curve for converting Stokes signals
to magnetic field strengths. Usually, the calibration applies
globally so that no pixel to pixel variation in the thermodynamic
variables is taken into account. Such procedures were chosen
by the authors of the first and sixth to eighth row
of Table~\ref{tab:angular_distribution}. The synthetic data of rows 11
and 12 were inverted with \citeauthor{ost_lites+al2008}'s method.
Of course, also this
second, more fundamental way of deriving the magnetic field vector
can be considered a Stokes inversion procedure---one is
tempted to call it a ``poor man's inversion procedure''. Its
advantage is that it immediately derives from the observed data 
and its model dependency is less complex.

One method on this second route for retrieving the 
true magnetic field strength of quiet Sun regions consists in 
considering the Stokes-V amplitude ratio of two spectral lines.
Traditionally, this was done with the two lines
\ion{Fe}{i} $\lambda\lambda$ 525.022 and 524.706~nm, which
yield the so called magnetic line ratio 
\citep{ost_stenflo1973}. The two lines have effective Land\'e factors 
3.0 and 2.0, respectively, but are otherwise identical in terms of 
line strength and excitation potential and belong to the same 
atomic multiplet. Therefore, the two lines behave virtually identical 
with regard to formation height and thermal response but they vary 
in the sensitivity to magnetic fields. 

In the weak field regime in which the Zeeman splitting is much
smaller than the width of the spectral line (at most a few hundred
Gauss), 
\begin{equation}
\label{eq:vpropto}
  V(\lambda) \propto g\frac{\partial I(\lambda)}{\partial \lambda}\;,
\end{equation}
where $V(\lambda)$ is the Stokes $V$ profile, the difference between 
left and right circularly polarized intensity  
and $I(\lambda)$ the Stokes $I$ profile, the regular, natural light 
intensity (absorption profile). $g$ is the effective Land\'e factor. 
Equation~(\ref{eq:vpropto}) simply derives from the leading term
of the Taylor expansion of the two Zeeman components expressed
in terms of the $I$ profile 
\citep[see e.g.][chapter 12.2]{ost_stenflo1994}. Then the
Stokes $V$ ratio of two lines 1 and 2 is
\begin{equation}
\label{eq:ratio}
  \frac{V_1}{V_2} = \frac{g_1}{g_2}
  \frac{{\partial I_1}/{\partial \lambda}}{{\partial I_2}/
  {\partial \lambda}}\approx\frac{g_1}{g_2}\;,
\end{equation}
which is in particular also valid for the wavelength position where
Stokes $V$ assumes its peak value.
The second relation of Eq.~(\ref{eq:ratio}) is well satisfied only
when lines 1 and 2 are virtually identical and form under very
similar conditions as is the case for the line pair  
\ion{Fe}{i} $\lambda\lambda$ 525.022 and 524.706~nm.
As long as the magnetic field is intrinsically weak,
the Stokes-$V$ amplitude-ratio is $g_1/g_2$. But with increasing
field strength, the Stokes-$V$ amplitude starts to saturate,
first the line with the larger $g$-factor while the $V$ amplitude 
of the line with the smaller $g$-factor continues to grow and 
Eq.~(\ref{eq:ratio}) breaks down because higher order terms in
the above mentioned Taylor expansion come into play. At
field strength in the order of 1~kG, the two amplitudes are of
similar size, meaning that the amplitude ratio approaches 
$V_1/V_2\approx 1$. Further increase of the magnetic field
strength merely increases the Zeeman splitting between the blue
and the red lobe of the Stokes profile, linearly with field strength. 
Thus, the line-ratio method provides an easy and comprehensible 
tool for differentiating
between weak and strong fields in the regime where full
Zeeman splitting of the lines has not yet taken place, making
it possible to dispense with detailed radiation transfer and 
modeling. It is important to notice that the line-ratio method
also works when only a small fraction of the spatial resolution 
element is occupied by magnetic fields (small filling factor)
so that the method yields information about sub-resolution
magnetic fields, i.e., the \emph{intrinsic} field strength.

\begin{figure}[t]
\centerline{\includegraphics[width=0.75\textwidth]{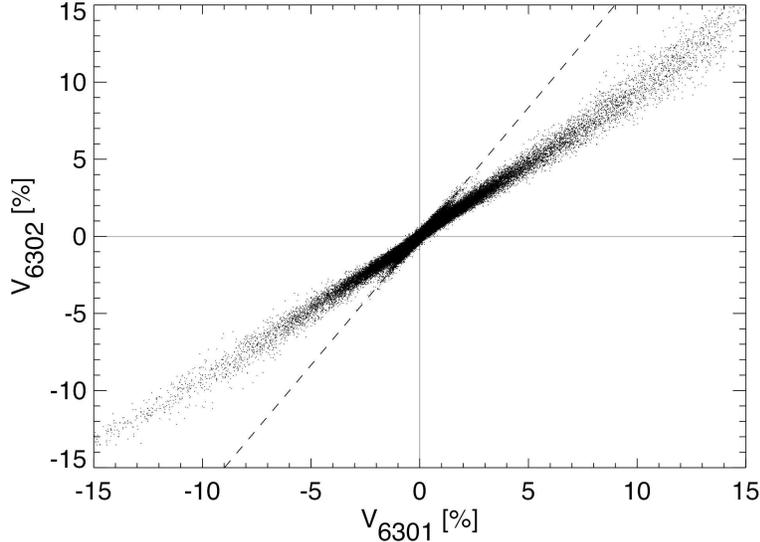}}
\caption{Scatter plot of the blue lobe Stokes-$V$ amplitude of
\ion{Fe}{i} 630.250~nm vs.\ the corresponding amplitude of \ion{Fe}{i} 630.151~nm
from the leftmost third of the `normal mode' map shown in Figs.~1 and 2 of 
\citet{ost_lites+al2008}.
The dashed line with slope $s=1.66$ represents the regression relation that 
would be expected for intrinsically weak fields. Note the two populations of 
points: (1) The dominant population that follows a slope of about 
$0.65 s$ for small polarizations. 
(2) A secondary population of Stokes $V$ amplitudes 
below about 2\%, which closely follows the dashed line. The method for
determining the Stokes-$V$ amplitudes was the same as used by
\citet{ost_stenflo2010} (Fig.~8) for a different (`deep mode') data set. 
\label{fig:scat_stenflo}}
\end{figure}

Unfortunately, Hinode SOT/SP does not detect the ideal line pair 
\ion{Fe}{i} $\lambda\lambda$ 525.022 and 524.706~nm but instead 
the line pair
\ion{Fe}{i} $\lambda\lambda$ 630.151 and 630.250~nm with respective
Land\'e factors 1.667 and 2.5, for which the second relation of 
Eq.~(\ref{eq:ratio}) is less well satisfied because the ratio
of Stokes-$I$ derivatives does not cancel out but introduces
an intricate dependence on temperature and velocity in the
atmosphere \citep{ost_khomenko+collados2007}. Therefore it did not seem 
advisable to apply the line-ratio method to SOT/SP data. Nevertheless, 
\citet{ost_stenflo2010} gave it a try, which led him to a remarkable discovery 
which is displayed in Fig.~\ref{fig:scat_stenflo}.

Fig.~\ref{fig:scat_stenflo} shows a scatter plot of the blue lobe Stokes-$V$ amplitude of
\ion{Fe}{i} 630.250~nm vs.\ the corresponding amplitude of \ion{Fe}{i} 630.151~nm
for a SOT/SP ``normal mode'' data set recorded at quiet Sun disk-center, viz., the
leftmost 100\arcsec\ of the map shown in Figs.~1 and 2 in \citet{ost_lites+al2008}.
Fig.~\ref{fig:scat_stenflo} is from a different dataset than, but very similar to 
Fig.~8 of \citet{ost_stenflo2010} and the Stokes-$V$ amplitudes were determined with
the same method as described in \citet{ost_stenflo2010}. 
The dashed line with slope $s=1.66$ represents the regression relation that 
would be expected for intrinsically weak fields. It is close to the ratio
of the respective Land\'e factors, $g_{630.25}/g_{630.15} = 2.5/1.667 = 1.5$  
but not exactly because the ratio of the derivatives in Eq.~(\ref{eq:ratio}) 
is not unity but rather 1.1 at the wavelength position of maximum 
Stokes $V$ amplitude of the blue lobe. The unexpected result is that
there exists two populations of points: (1) The dominant population that follows 
a slope that is about $0.65 s$ for small polarizations, decreasing to about $0.55 s$ 
for larger values. (2) A secondary population that is significant only for 
Stokes $V$ amplitudes below about 2\%, and which closely follows the dashed line.
Even at the 1\% polarization level, the two components can be clearly separated
when plotting the histogram of the points in a bin around this polarization 
level \citep[][Fig.~9]{ost_stenflo2010}. Reliable decompositions are found down
to $V_{630.15} \ga 0.6$\%.

Fig.~\ref{fig:scat_stenflo} has a simple and elegant explanation in terms of
the Stokes-$V$ line-ratio theory. Clearly, the points of the second population
must correspond to intrinsically weak fields because their Stokes-$V$ amplitude
ratios are in close agreement with the weak field limit (dashed regression
relation). The points of the first population must correspond to kG fields because 
their Stokes-$V$ amplitude ratios is approaching, and even dropping below,
one, which points to strong fields as explained above. 
But how is it possible that the strong field component exists not only for 
large polarization amplitudes, which may stem from network elements, but extends 
down to very small polarization levels?
One would rather expect the scatter plot to display a S-shape---the inner part
being occupied by the weak field component with slope $s\approx 1.66$ and the outer 
part by the strong field component with slope $s\approx 1.1$. The answer to
this question is filling factor. There must exist ample amounts of kG-fields 
of sub-resolution size (small magnetic filling factor), which give only weak 
polarization amplitudes but nevertheless betray themselves as being intrinsically 
strong because of a small Stokes-$V$ amplitude ratio. Thus, pixels with polarization
levels around and below 1\% must harbor kG flux concentrations of very small size
and in fact, \citet{ost_stenflo2011} deduces from these results kG flux tubes with diameters 
down to 10~km. As elegant and obvious this 
explanation of Fig.~\ref{fig:scat_stenflo} may appear, caution is indicated as is 
demonstrated in the following.

\begin{figure}[t]
\centerline{\includegraphics[width=0.48\textwidth]{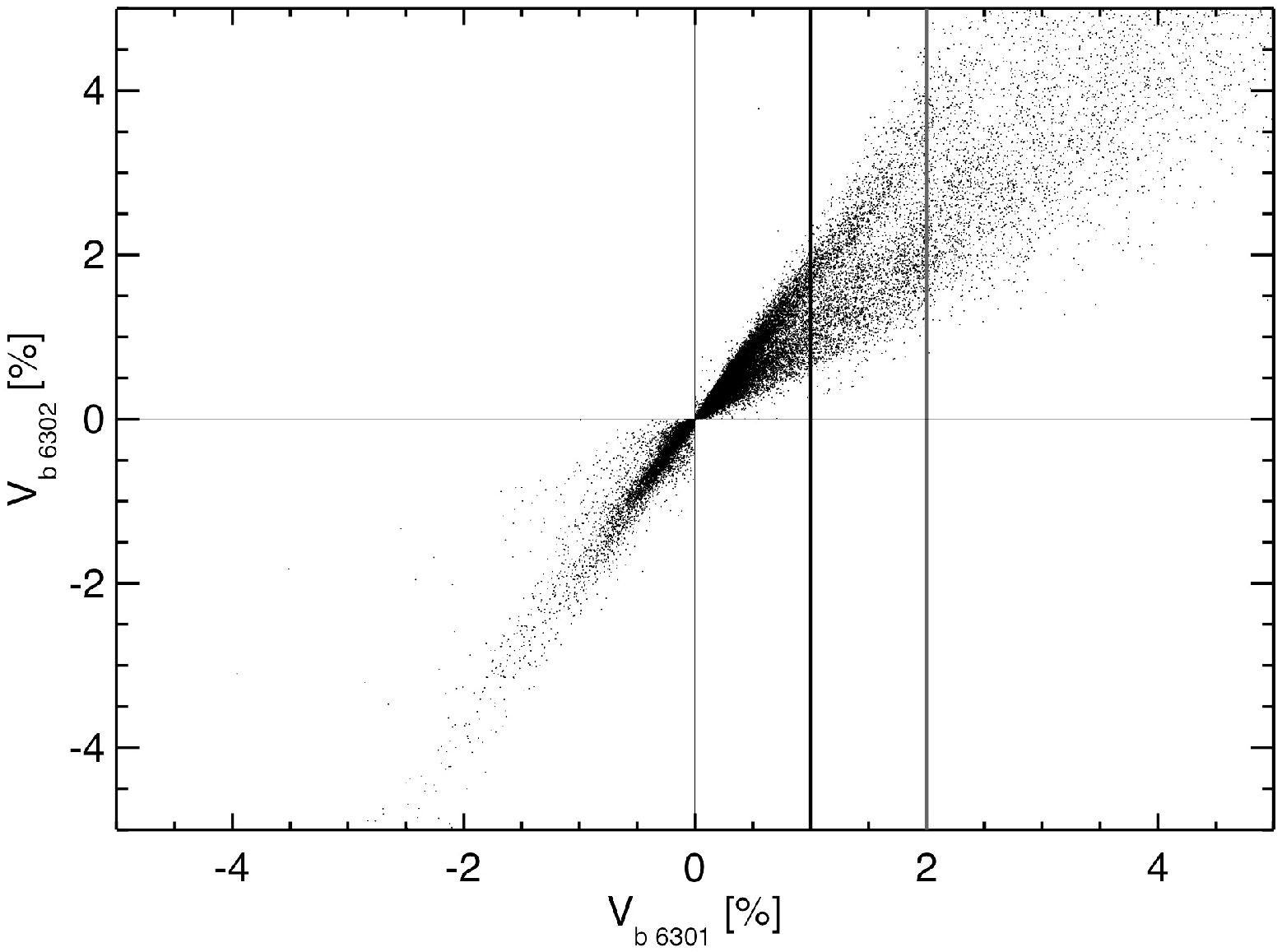}\hspace*{\fill}
            \includegraphics[width=0.48\textwidth]{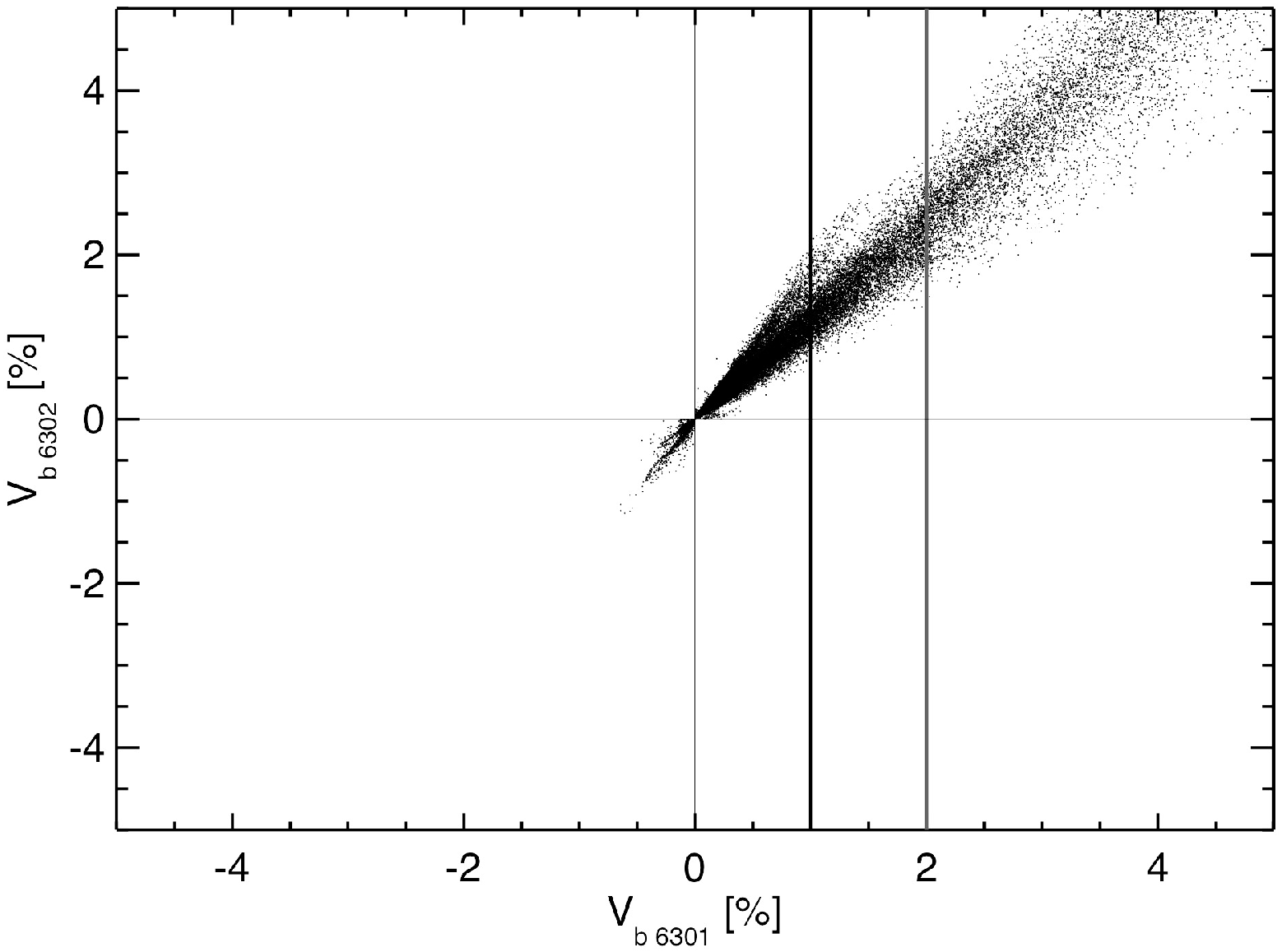}}\hspace*{\fill}
\centerline{\includegraphics[width=0.48\textwidth]{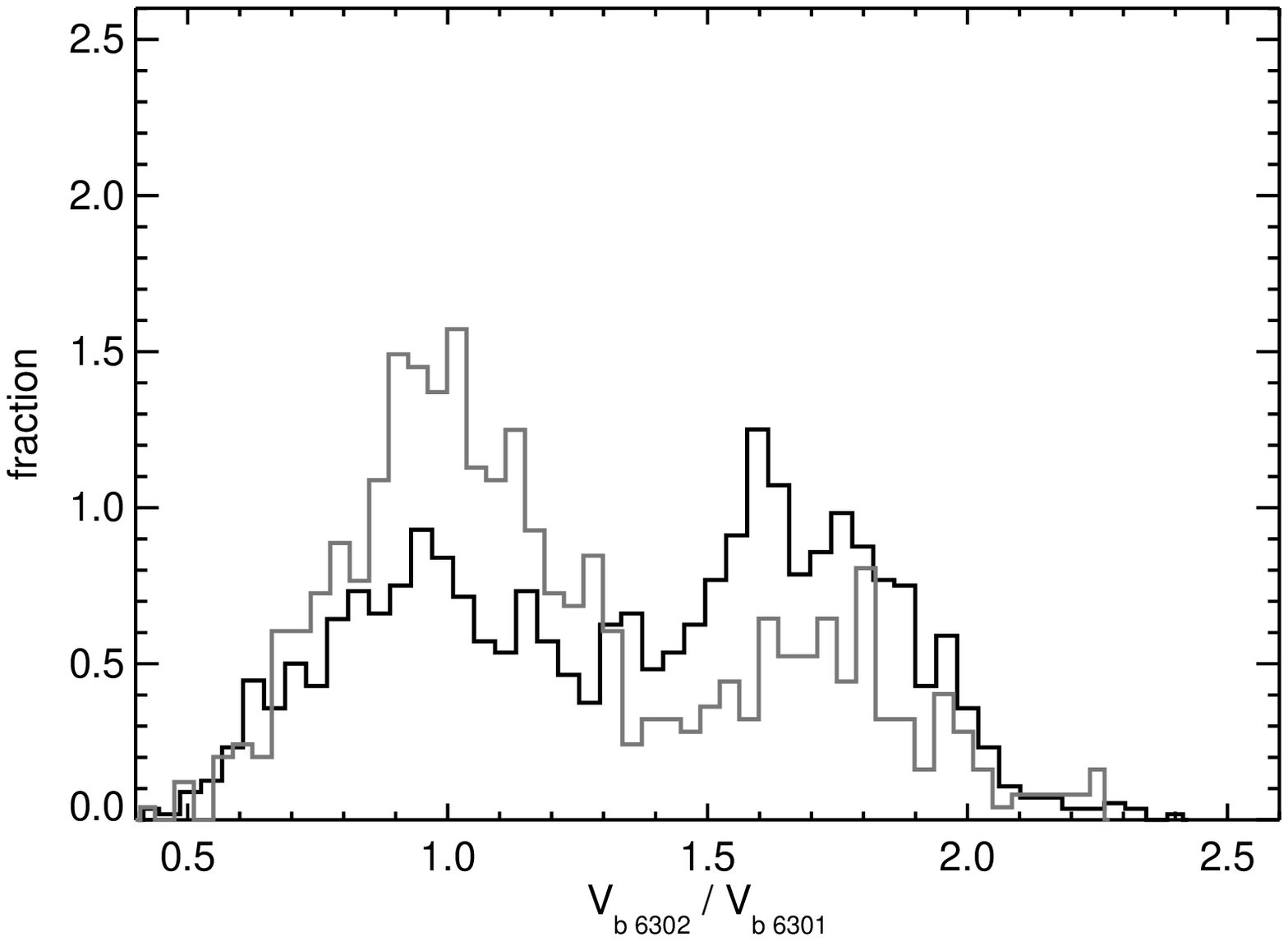}\hspace*{\fill}
            \includegraphics[width=0.48\textwidth]{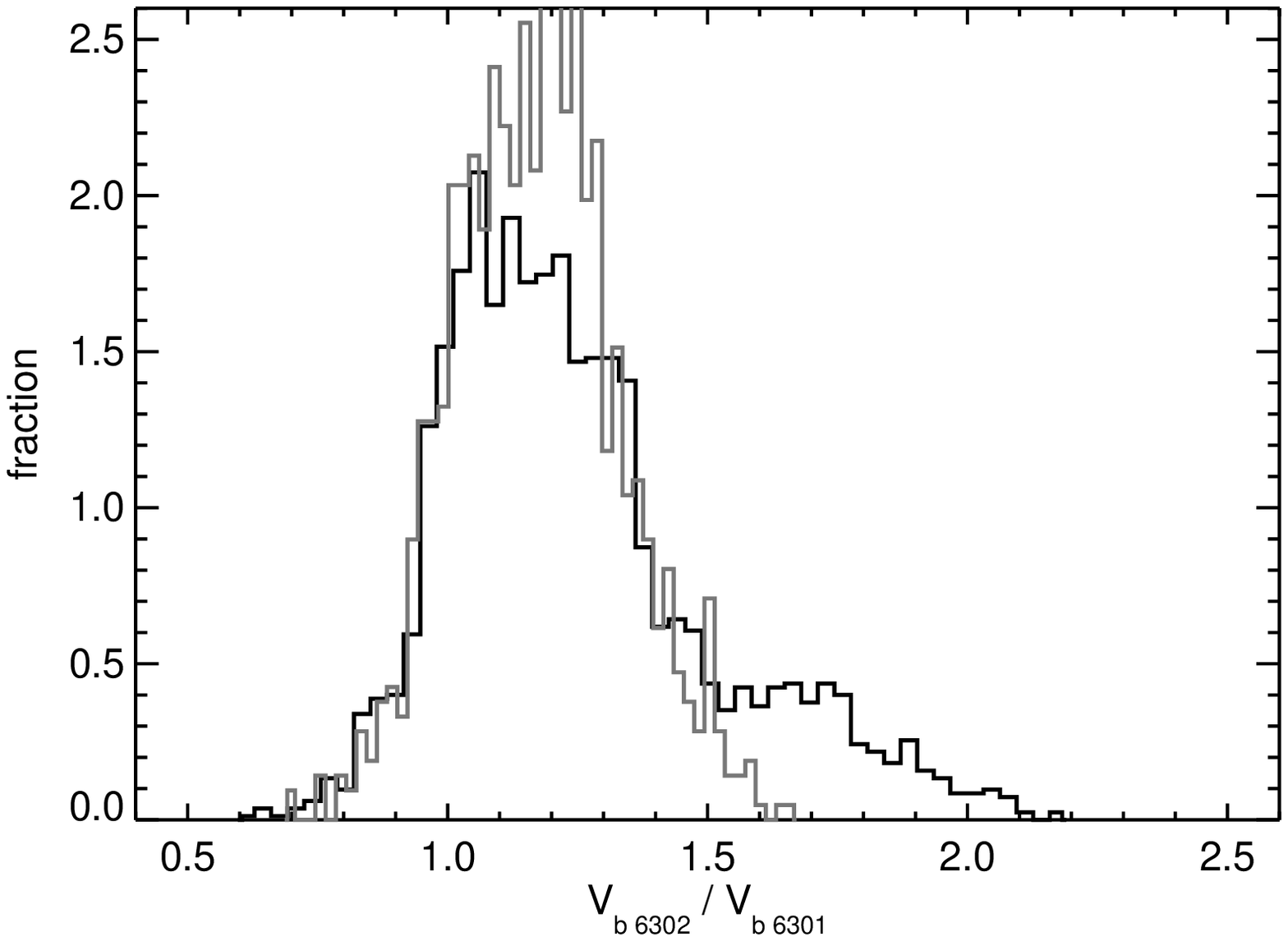}}\hspace*{\fill}
\caption{Top row: Scatter plot of the Stokes-$V$ amplitude of the blue lobe of
\ion{Fe}{i} 630.250~nm vs.\ the corresponding amplitude of \ion{Fe}{i} 630.151~nm
of synthesized $V$ profiles from the CO5BOLD MHD simulation v50. The scatter to the 
left is from pixels at full spatial resolution,  the scatter to the right from the 
same data after application of the SOT PSF. The vertical lines indicate the polarization
levels which the histograms in the bottom row correspond to.
Bottom row: Histograms of the points that fall within the range 
$0.8 \le V_{630.15} \le 1.0$ (black) and 
$1.8 \le V_{630.15} \le 2.0$ (gray).
\label{fig:scat_v50}}
\end{figure}

\subsection{Stokes-{\boldmath $V$} Amplitude Ratios from Simulations}
\label{sec:line_ratio_sim}
Fig.~\ref{fig:scat_v50} shows the inner part of a scatter plot of the same
quantities as shown in Fig.~\ref{fig:scat_stenflo} but of Stokes-$V$ profiles
that were synthesized from the CO5BOLD MHD-simulaion v50 referred to in
Sect.~\ref{sec:numerical_simulations}. The top left panel shows the scatter
plot from the data at full spatial resolution, the top right panel  
after application of the Hinode/SOT point spread function (PSF) of \citet{ost_wedemeyer2008} 
to the full resolution data. Here, we do not use Stenflo's fitting method for
determining the Stokes-$V$ amplitudes but take the actual amplitude, discarding
abnormal Stokes-$V$ profiles. The bottom panels show the corresponding histograms
of the points that fall within the range $0.8 \le V_{630.15} \le 1.0$ (black) and 
$1.8 \le V_{630.15} \le 2.0$ (gray), again, once from the full resolution data
(left) and once from the data after application of the SOT PSF (right).
We see immediately that, like in the real, observed data, there are
two populations of points: (1) The dominant population that follows 
a slope that is about $0.6 s$. (2) A secondary population that is significant only for 
Stokes $V$ amplitudes below roughly 2\%, and which approximately follows a
regression line with slope $s=1.66$. For the full resolution data, even at 
the 1\% polarization level, the two components can be clearly separated  
as can be seen from the two distinct humps in the corresponding histogram.

\begin{figure}[h]
\centerline{\includegraphics[width=0.48\textwidth]{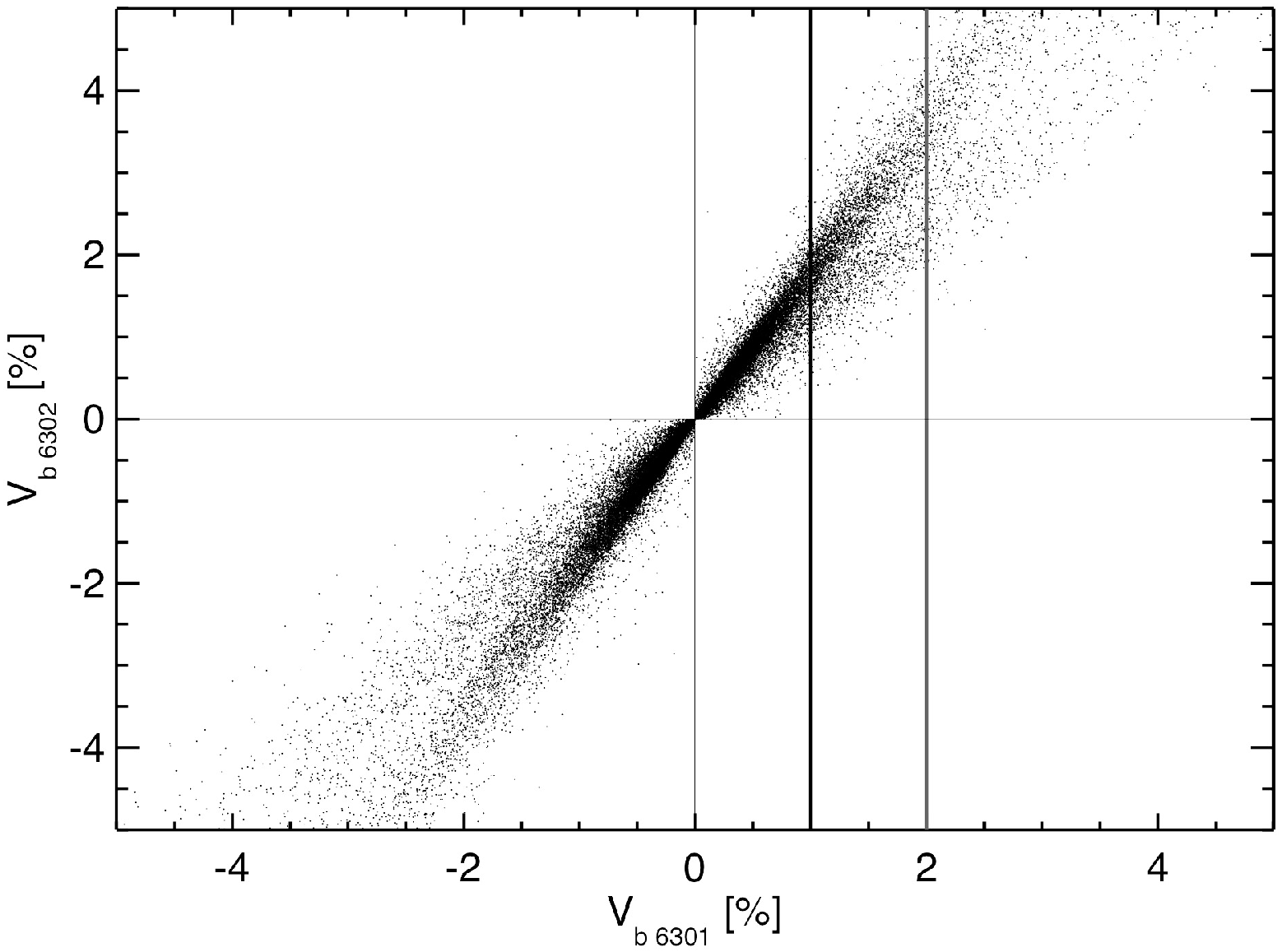}\hspace{\fill}
            \includegraphics[width=0.48\textwidth]{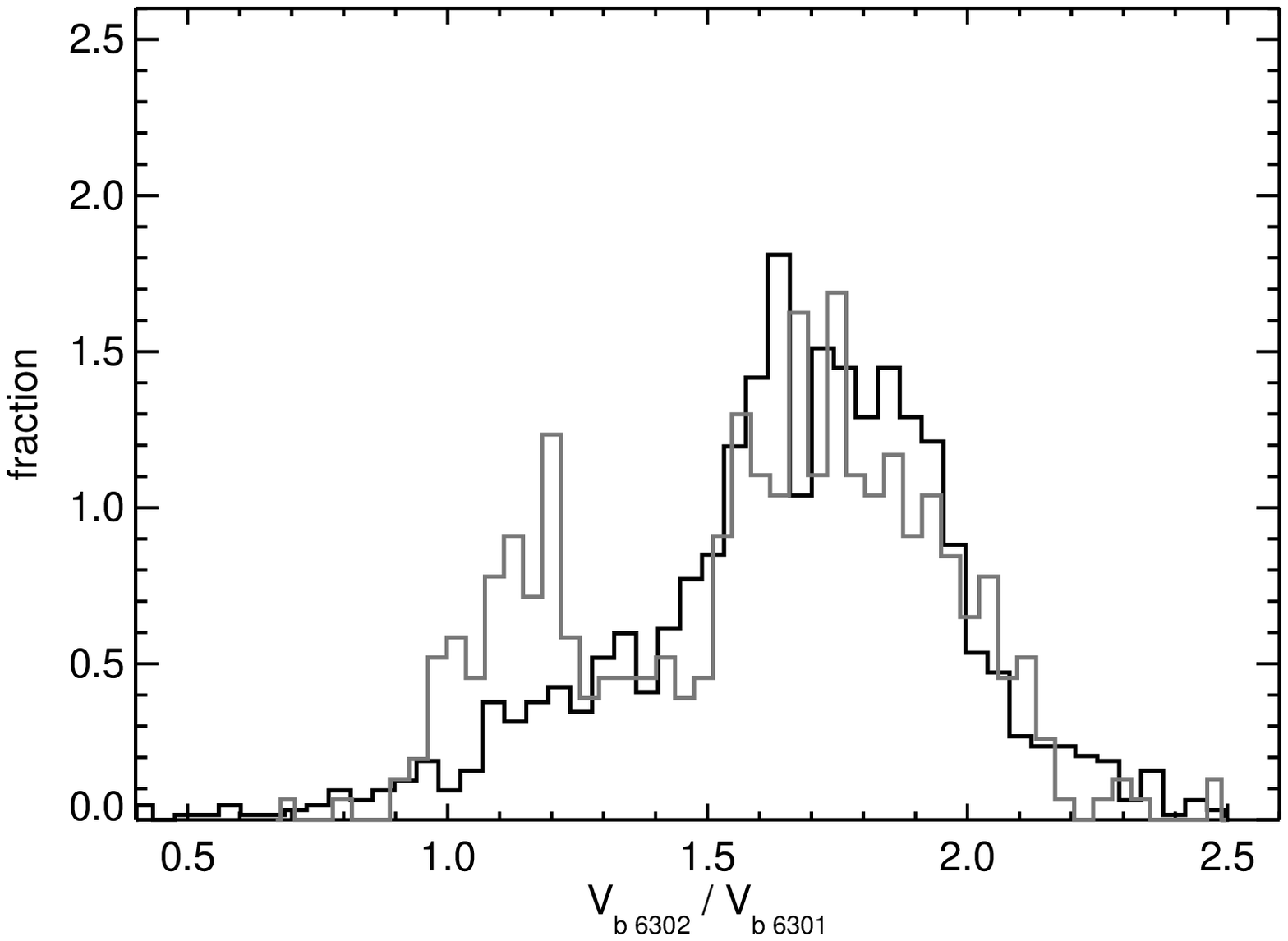}}
\caption{Left: Scatter plot of the blue lobe Stokes-$V$ amplitude of
\ion{Fe}{i} 630.250~nm vs.\ the corresponding amplitude of \ion{Fe}{i} 630.151~nm
of synthesized $V$ profiles from the CO5BOLD MHD simulation h50. The scatter is from 
pixels at full spatial resolution. 
Right: Corresponding histogram of the points that fall within the range 
$0.8 \le V_{630.15} \le 1.0$ (black) and $1.8 \le V_{630.15} \le 2.0$ (gray).
\label{fig:scat_h50}}
\end{figure}

A scatter plot from a snapshot of the simulation h50 (Fig.~\ref{fig:scat_h50})
shows a similar behavior with the difference that both polarities are equally
represented and that the weak field population is now the major one. This is
because in this simulation the field is weaker and more turbulent than in simulation 
v50 and has no preferred polarity from the beginning.

On the background of the explanation for the observed scatter plot in terms of
intrinsically weak and strong fields and in terms of the magnetic filling factor, 
the result from the synthesized data is very surprising. Since the simulations 
cannot harbor any sub-resolution magnetic elements because there cannot
be any structure smaller than the resolution limit given by the computational 
grid, the magnetic area filling factor for a single computational cell is always 
one, irrespective of the field strength. Thus, the argument that pixels of small 
polarization degree with a Stokes-$V$ line-ratio in the order of unity must 
origin from kG flux concentrations of small filling factor cannot apply to the 
simulation data. For those, the two populations of points in the scatter plot 
must have a different origin.

\subsection{Interpretation of the Simulation Data}
\label{sec:interpretation_sim}
On the one hand side, it is reassuring that the MHD simulations reproduce
the observed two populations of points in scatter plots of the Stokes-$V$
amplitudes of the lines \ion{Fe}{i} $\lambda\lambda$ 630.151 and 630.250~nm.
On the other hand, we found in Sect.~\ref{sec:line_ratio_sim} that the
two populations cannot be explained in terms of the very attractive and
elegant hypothesis put forward by \citet{ost_stenflo2010} of a ``magnetic dichotomy'' 
with two distinct populations representing strong (kG) and weak fields, or 
``collapsed'' and ``uncollapsed'' fields as he expresses it. The question then is, 
what causes the two populations if not the dichotomy of ``collapsed'' and 
``uncollapsed'' fields?

\begin{figure}[h]
\centerline{\includegraphics[width=0.9\textwidth]{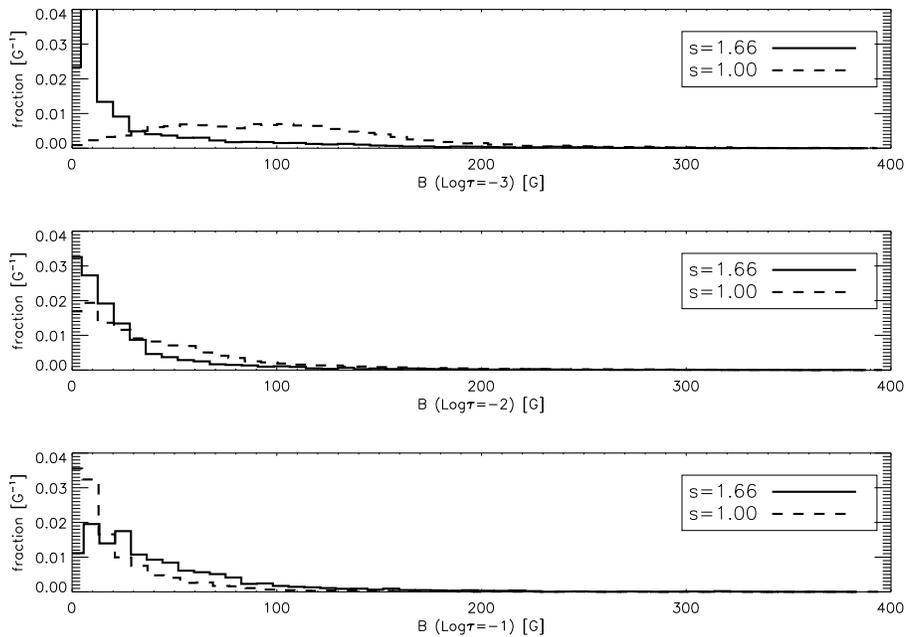}}
\caption{Histograms of the absolute magnetic field strength at three different
optical depth levels $\log\tau = -3$ (top), $-2$ (middle), and $-1$ (bottom), 
for the points of the full resolution scatter plot of Fig.~\ref{fig:scat_v50} 
with a polarization level of $1.0 \le V_{630.15} \le 2.0$. 
Dashed curve: Histograms of the main population of points with Stokes-$V$ amplitude 
ration $V_{630.25}^{\rm b}/V_{630.15}^{\rm b} \le 1.5$.
Solid curve: Histogram of the population with Stokes-$V$ amplitude ration 
$V_{630.25}^{\rm b}/V_{630.15}^{\rm b} \ge 1.5$.
\label{fig:B_histogram}}
\end{figure}

\begin{figure}[]
\centerline{\includegraphics[width=0.7\textwidth]{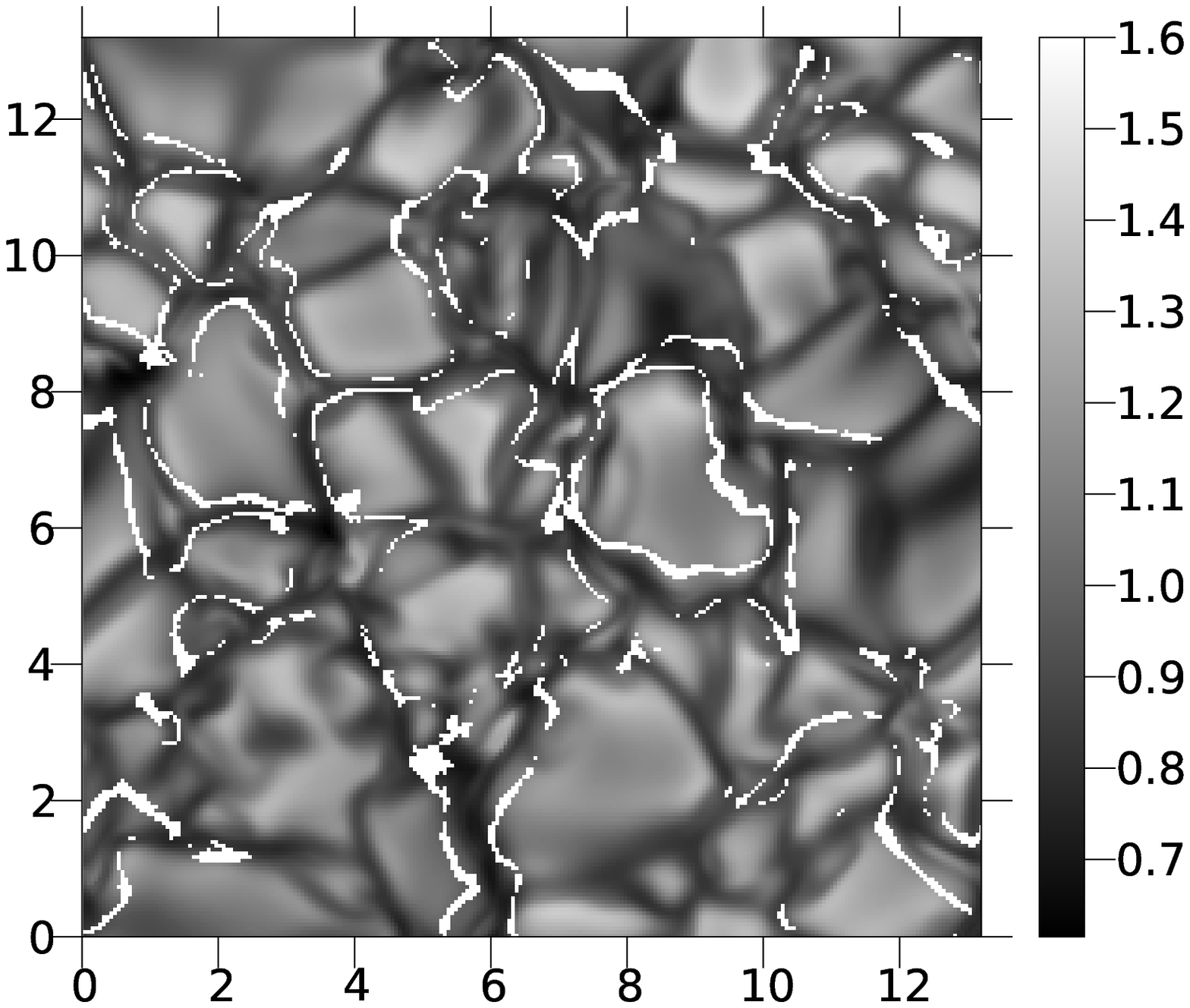}}
\vspace*{-0.3em}
\centerline{\includegraphics[width=0.7\textwidth]{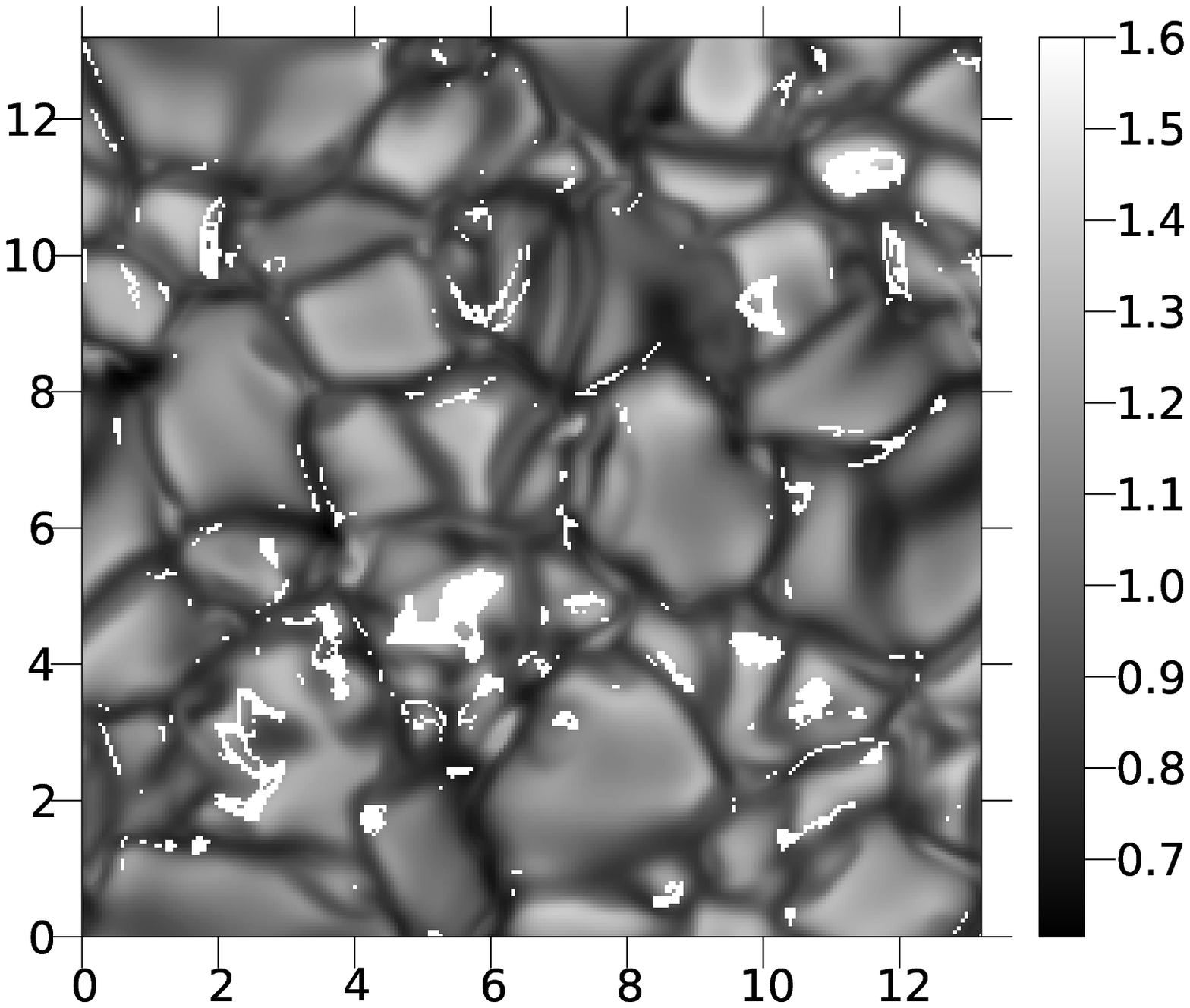}}
\caption{Top: Location of pixels (white) with a polarization level 
$1.0 \le V_{630.15} \le 2.0$ belonging to the population of pixels
with $V_{630.25}^{\rm b}/V_{630.15}^{\rm b} \le 1.5$ (first, main
population). Bottom: Location of pixels with a polarization level 
$1.0 \le V_{630.15} \le 2.0$ belonging to the population
with $V_{630.25}^{\rm b}/V_{630.15}^{\rm b} \ge 1.5$ (second, intrinsic
weak field population). Background: Continuum intensity at 630~nm.
\label{fig:population+continuum}}
\end{figure}

Figure~\ref{fig:B_histogram} partially answers this question. It shows
the histograms of the absolute magnetic field strength at three different
optical depth levels $\log\tau = -3$ to $-1$ for the points of the 
full resolution scatter-plot of Fig.~\ref{fig:scat_v50} with 
a polarization level of $1.0 \le V_{630.15} \le 2.0$. 
Each panel contains two histograms: one for the main population of points 
with Stokes-$V$ line-ration 
$V_{630.25}^{\rm b}/V_{630.15}^{\rm b} \le 1.5$ (dashed) and
one for the population of points with
Stokes-$V$ line-ration 
$V_{630.25}^{\rm b}/V_{630.15}^{\rm b} \ge 1.5$ (solid).
The second population should correspond to intrinsically weak magnetic
fields because they cluster around  a Stokes-$V$ amplitude ratio of 
1.6, which is the value in agreement with the weak field approximation.
In fact, we can see from the histograms that this is true: the field
strength for this population stays essentially below 200~G with
a peak at 10~G. But we also see that the first population
that should correspond to ``collapsed'' kG flux concentrations is
weak field as well with field strengths mostly below 300~G. There
is however a distinct difference with respect to the second population:
the peak and bulk of the histogram shifts to higher field strengths with
height in the atmosphere. This means that for this population of points,
the magnetic field strength increases with height in the atmosphere, which 
is opposite to the usual case.

When plotting the histogram of the magnetic field inclination with 
respect to the vertical direction (not shown here) we see that the
first population shows mainly horizontal fields in the deep layers, 
which  become inclined by $\approx 45\deg$ at 
$\log\tau = -2 $ to $-3$ when the field strength increases. 
This behavior points towards canopy magnetic fields as the origin
of this, formerly called, the ``collapsed''
\citep{ost_stenflo2010} component. Canopy fields may stem from magnetic 
flux concentrations in the photosphere, which expand and fan out in 
the lateral direction with increasing height in the atmosphere 
\citep{ost_steiner_encyc2000}. The magnetic field along a line of sight 
that traverses the canopy field is weak in the optically deep layers 
below the canopy but abruptly  increases in strength and assumes a
large inclination angle at the transition from below to within the 
canopy. Another indication that points to canopy fields comes from
the histograms of the vertical velocity, which show mainly upflows 
at all optical depth levels for the second, intrinsically weak field
population, but downflows in the deepest layer of the first population.
This can be expected because canopy fields are most likely found in the
vicinity of (vertical) magnetic flux concentrations which in turn are found
within intergranular lanes, which harbor downflows. Therefore, we expect
canopy fields and hence, pixels belonging to the first population, 
to exist at the border between intergranular lanes and 
granules with a tendency of having downflows in the deep layers.

Figure~\ref{fig:population+continuum} confirms that this is indeed the case.
The top panel shows on the background of the continuum intensity at 630~nm the 
location of pixels that have a polarization level $1.0 \le V_{630.15} \le 2.0$ 
and belong to the first population with 
$V_{630.25}^{\rm b}/V_{630.15}^{\rm b} \le 1.5$. 
Clearly, these pixels are found along the boundaries of granules, where canopy 
fields exist. The bottom panel shows the pixels belonging to the second 
population with $V_{630.25}^{\rm b}/V_{630.15}^{\rm b} \ge 1.5$.
These pixels occur in patches within granules and are mainly associated with
horizontal field, which is probably transported into the 
photosphere by instances of granular overshooting.

From all this we conclude that the two population of pixels with low
polarization amplitudes found in data of synthesized Stokes-$V$ profiles
of magneto-convection simulations are \emph{not} due to a dichotomy between 
``collapsed'' and ``uncollapsed'' fields, but to a distinct difference
between profiles from granule interiors and profiles from the boundary
region between granules and the intergranular space. Truly strong fields
are found within the intergranular lane proper but they have a larger
polarization degree and therefore pertain to the outskirts of
the line-ratio scatter-plots. It is very likely, that this explanation
of the two populations also holds for the observed Stokes-$V$ profiles.
This then means that there is no need for the introduction of collapse
sub-resolution flux-concentrations, and that the conclusions of
\citet{ost_stenflo2010} that the inter-network magnetic field was 
predominantly vertical is premature because of its shaky premises.

\subsection{The Horizontal Field from Numerical Simulations}
\label{sec:horizontal_sim}
From the critiques reviewed in Sects.~\ref{sec:critiques} and \ref{sec:photon_noise}
it is clear that the angular distribution of the magnetic field in network-cell
interiors has not yet been reliably measured. It is not even clear if there
exists indeed a predominance of the horizontal field component. The analysis
of \citet{ost_lites+al2007,ost_lites+al2008}, which uses and derives least noise affected
data, points in this direction. On the other hand, numerical MHD simulations
of the solar surface layers have resulted in a predominance of the horizontal fields
\citep{ost_grossmann+al1998,ost_abbett2007,ost_schuessler+voegler2008,ost_steiner+al2008,ost_danilovic+al2010} 
particularly in the higher layers of the photosphere and low chromosphere. Here, we corroborate
these findings with the simulation runs mentioned in Sect.~\ref{sec:numerical_simulations}.

\begin{figure}[h]
\centerline{\includegraphics[width=0.49\textwidth]{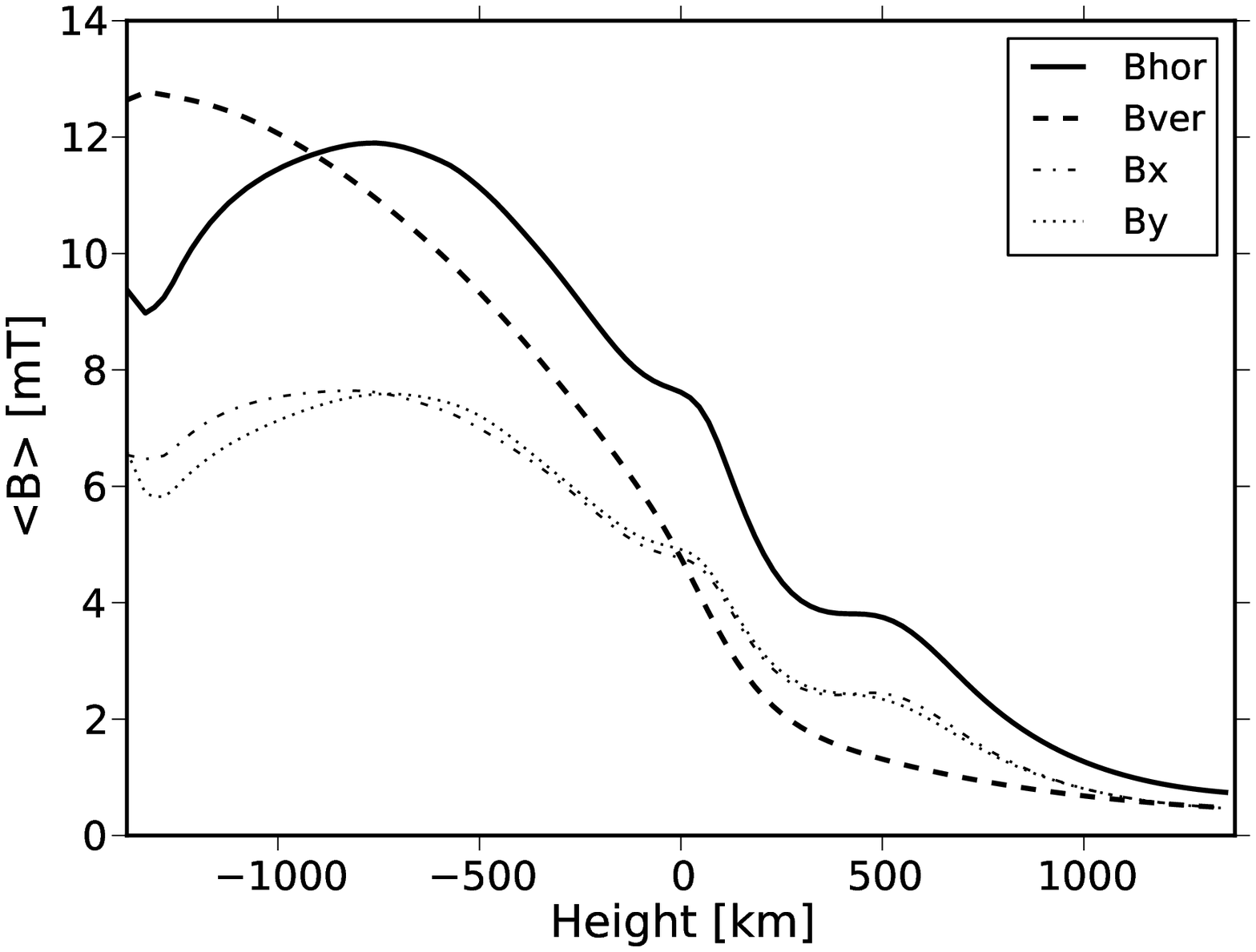}
            \hspace*{\fill}
            \includegraphics[width=0.475\textwidth]{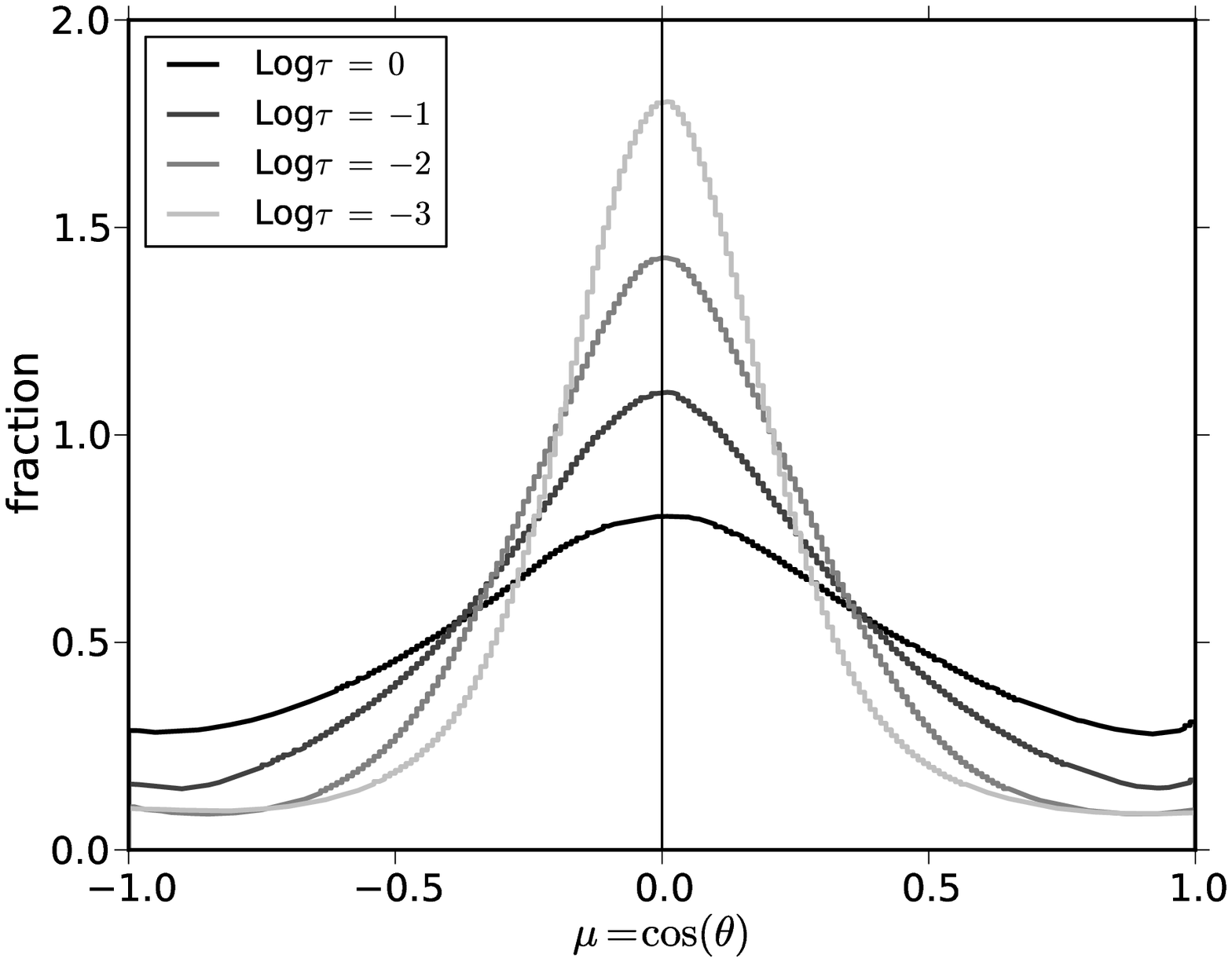}}
\caption{Left: Horizontally averaged strength of various magnetic field components
as a function of geometrical height in the model atmosphere of simulation h50. 
Dashed: Vertical component, $\langle |B_{\rm v}|\rangle$; Dot-dashed and dotted: 
Horizontal components $\langle |B_{x}|\rangle$ and $\langle |B_{y}|\rangle$; 
Solid: Horizontal component $B_{\rm h} = \langle (B_x^2 + B_Y^2)^{1/2}\rangle$.
Right: Histogram for the cosine of the inclination of the magnetic field with 
respect to the vertical direction for different optical depth levels from
$\log\tau = 0$ (broadest distribution), to $\log\tau = -3$ (narrowest distribution).
An isotropic distribution would give a straight horizontal line.
\label{fig:BhvsBv_h50}}
\end{figure}

Figure~\ref{fig:BhvsBv_h50} shows in the left panel the horizontally averaged
strength of various magnetic field component as a function of geometrical height 
in the model atmosphere of simulation h50. The dashed curve refers to the absolute 
value of the vertical component, $B_{\rm v}$, the dotted and dash-dotted curves to 
the absolute value of the two horizontal components $B_x$ and $B_y$ which stand 
orthogonal to each other, and the solid curve to the horizontal component 
$B_{\rm h} = (B_x^2 + B_Y^2)^{1/2}$.
The horizontal components surpass the vertical one throughout the photosphere
from $z=0$ to 500~km. They are particularly dominant at and closely above 
500~km height. At $z = 500$~km 
${\langle B_{\mathrm{h}}\rangle}/{\langle |B_{\mathrm{v}}|\rangle}(500\,\mbox{km}) = 2.9$
and 
${\langle B_{x, y}\rangle}/{\langle |B_{\mathrm{ver}}|\rangle}(500\,\mbox{km}) = 1.9$.

The panel on the right hand side of Fig.~\ref{fig:BhvsBv_h50} shows the histogram
for the cosine of the inclination of the magnetic field with respect to the vertical
direction for different optical depth levels from
$\log\tau = 0$ (broadest distribution), to $\log\tau = -3$ (narrowest distribution).
An isotropic distribution would be represented by a straight horizontal line.
Instead, the distributions are peaked at the $90\deg$ angle and become narrower
with increasing height in the atmosphere, because of the increasing dominance
of the horizontal component with height.

\begin{figure}[t]
\centerline{\includegraphics[width=0.52\textwidth]{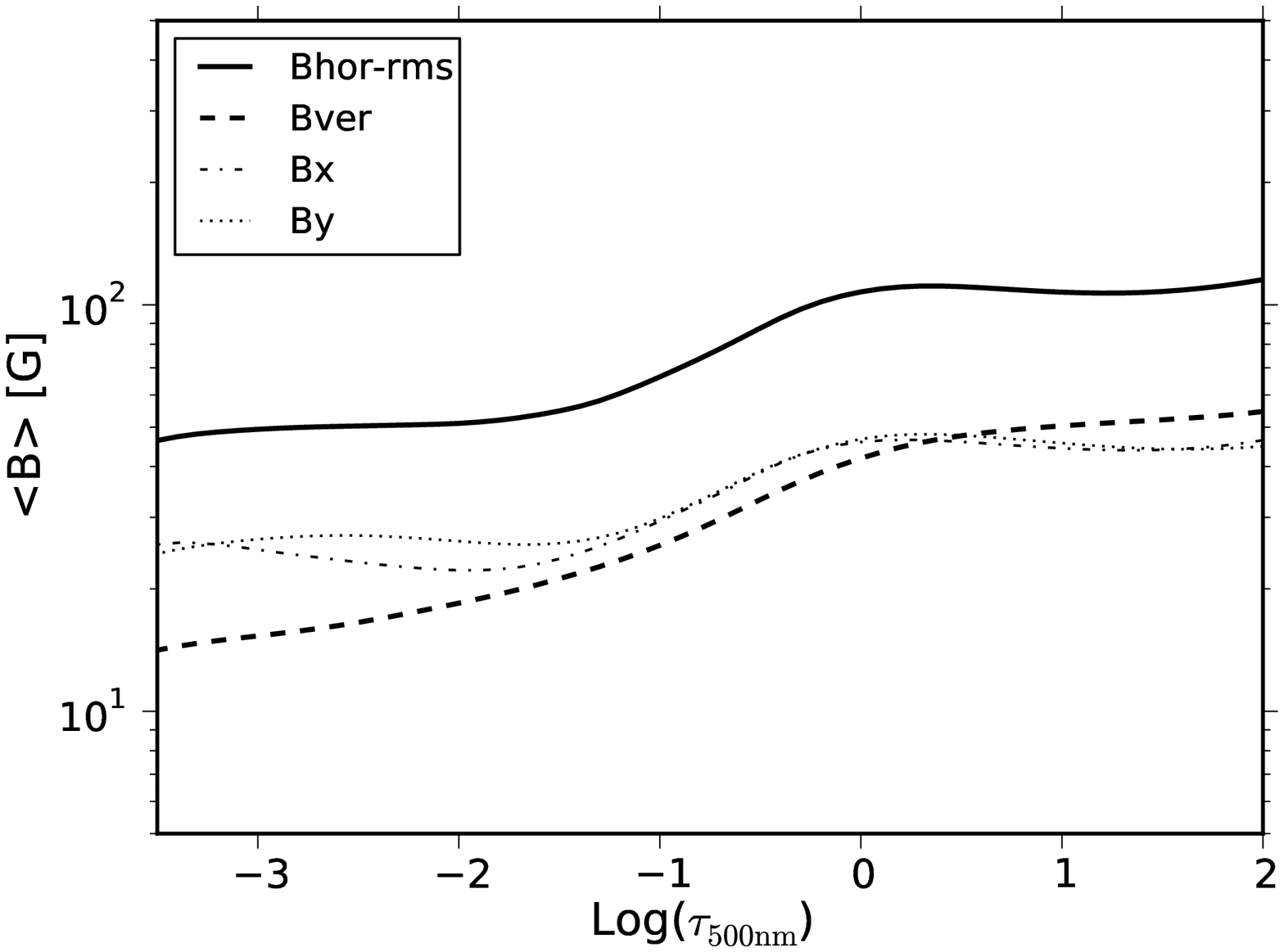}
\hspace*{\fill}\includegraphics[width=0.445\textwidth]{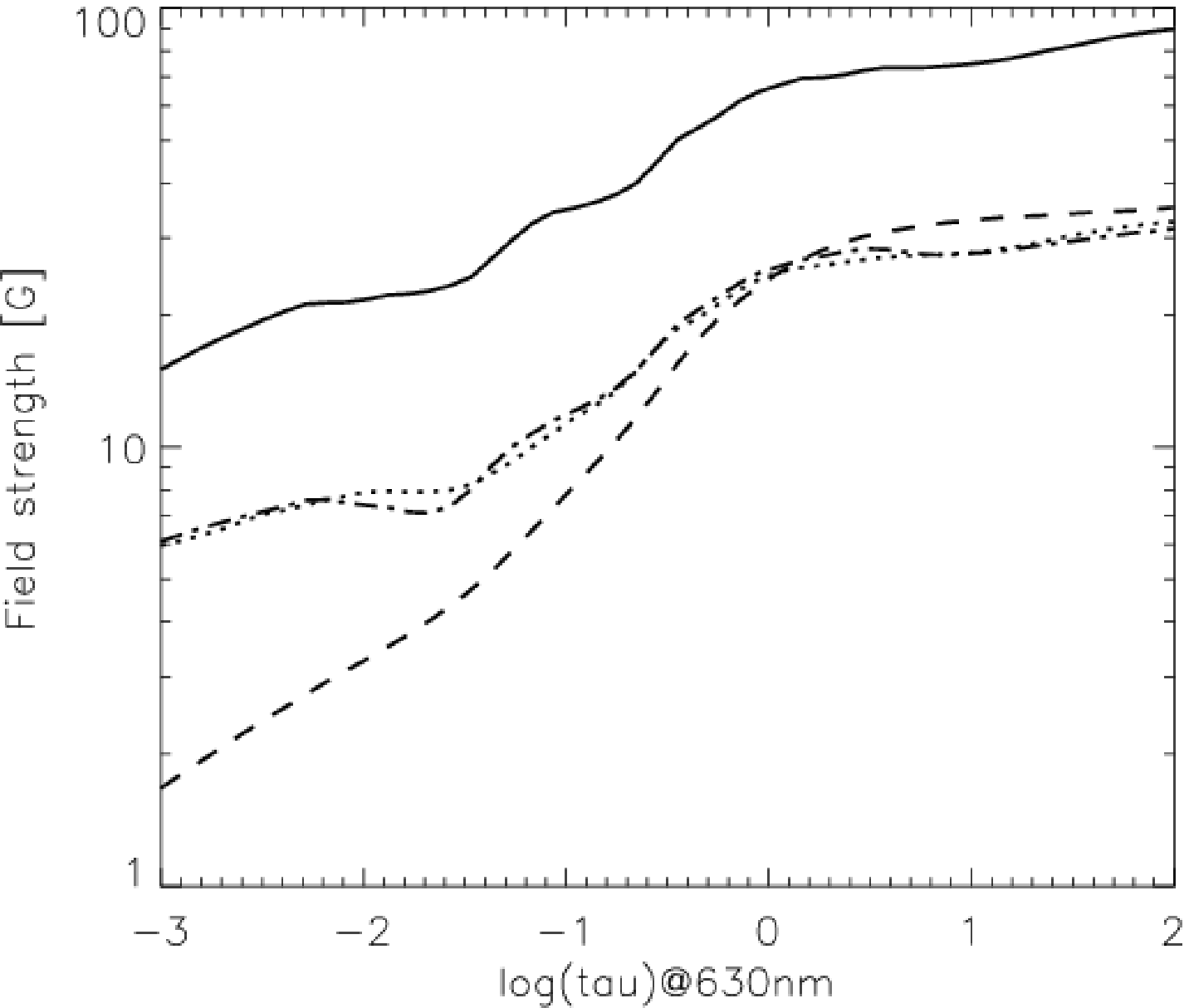}}
\caption{Horizontally averaged strength of various magnetic field components
as a function of optical depth of the model atmosphere of simulation h50 (left)
and the surface-dynamo simulation of \citet{ost_schuessler+voegler2008} (right, 
reproduced with permission \copyright\ ESO). 
Dashed: Vertical component, $\langle |B_{\rm v}|\rangle$; 
Dash-dotted and dotted: Horizontal components 
$\langle |B_{x}|\rangle$ and $\langle |B_{y}|\rangle$; 
Solid: Root-mean-square of the horizontal component 
$B_{\rm h, rms} = (\langle B_x^2 + B_Y^2)\rangle^{1/2}$.
\label{fig:comparison}}
\end{figure}

The situation is different for simulation v50. Even though, the horizontal components
show distinct local maxima close to $z = 500$~km, the vertical component dominates
for all heights. This is because of the initial condition of a homogeneous
vertical magnetic field $B_0$, which conserves the mean vertical field strength 
above the convection zone to $B_0$ for all times. Therefore, the ratio 
${\langle B_{\mathrm{h}}\rangle}/{\langle |B_{\mathrm{v}}|\rangle}$ is basically
predetermined by the initial condition in this case, which may be more representative
of a network-field patch or a weak plage region.

Figure~\ref{fig:comparison} shows a comparison between the simulations of 
\citet{ost_schuessler+voegler2008} (right panel) and the present result from 
simulation h50 (left panel). Again, the mean horizontal and vertical components 
are plotted, this time as a function of continuum optical depth and on a logarithmic 
scale. Despite the very different initial and boundary conditions and despite the
different codes
with which theses simulations were carried out, the general trend is similar.
In both simulations, the horizontal field component dominates throughout the 
photosphere. The absolute values of the field strength are different though:
in case of the simulation h50, it depends on the strength of the horizontal
field that is advected across the lower boundary (which is a boundary condition),
in case of \citet{ost_schuessler+voegler2008} it depends on the efficiency of the
surface dynamo, which in turn depends on the Reynolds numbers achieved by
the simulation.

\section{Summary and Conclusions}
Vortical flows of mainly vertically directed vorticity have been observed to exist 
in the deep photosphere 
\citep{ost_brandt+al1988,ost_wang+al1995,ost_bonet+al2008,ost_attie+al2009,%
ost_bonet+al2010,ost_balmaceda+al2010,ost_vargas-dominguez+al2011} and in the chromosphere
\citep{ost_wedemeyer+rouppe2009}. Numerical simulations show particularly strong vortical 
flows immediately below the visible optical surface $\tau_c = 1$, beneath 
(the vertices of) inter-granular lanes. The centripetal
force associated with these vortical flows is so strong that the gas pressure
gradient opposing it leads to a substantial reduction of the density within them, 
which may have observable consequences 
\citep{ost_nordlund1985,ost_kitiashvili+al2011,ost_moll+al2011}. We find that there exists no 
one-to-one connection between the density deficient vortices in the surface layers 
of the convection zone and the observable vortical flows at $\tau_c = 1$, but one can
assume the latter to extend and intensify with depth, simply because they are
found within downdrafts and because of angular momentum conservation in combination
with the gravitational stratification. This indicates that vortex stretching is
probably an important source of vorticity in the surface layers although 
\citet{ost_shelyag+al2011} found the baroclinic vorticity generation to dominate
in the convection zone and at the visible surface.

As was anticipated by \citet{ost_nordlund1985} and \citet{ost_brandt+al1988}, 
these vortical flows can be an important source of hydromagnetic disturbances.
Magnetic field that is advected into vortices start to rotate. In
the photosphere, where the thermal pressure generally dominates the magnetic 
energy density (\,$\beta >> 1$), this can be considered a kinematic process, but vertically
extending magnetic fields quickly start to dominate with height in the atmosphere.
When $\beta << 1$, the rotating magnetic field now acts on the plasma via Lorentz force
making it to rotate. Therefore, in the photosphere and in the chromosphere,
the surface of $\beta=1$ separates two regimes of vorticity generation
(see also Fig.~\ref{fig:swirl_co5bold3d}). In the regime, where $\beta << 1$
vortical flows of vertically directed vorticity is generated by vertically
extending magnetic flux concentrations which root within vortical flows of
the deep photosphere and the surface layers of the convection zone. Thus,
while vorticity is generated by hydrodynamic processes (vortex stretching and
baroclinic vortex generation) in the deep photosphere and the convection
zone, it is mediated to the upper photosphere and chromosphere via magnetic
fields. One can imagine that the formation of rotating magnetic flux 
concentrations may be the source of torsional Alfv\'en waves \citep{ost_jess+al2009}, 
which could potentially create a substantial Poynting flux in the outward direction, 
leading to coronal disturbances (but see \citep{ost_routh+al2010} for the cutoff
frequency of torsional tube waves) and having consequences for solar-wind acceleration 
and coronal heating in regions of open magnetic field configurations \citep{ost_zirker1993}. 
 
Granular lanes are composed of a leading bright rim and a trailing dark edge, which 
form at the boundary of a granule and move together into the granule itself. Virtually
every granule harbors one or several granular lanes during its lifetime and they
are well visible in high resolution broad band filtergrams of the solar surface.
They are the visible manifestation of horizontally extending vortex tube
found in numerical simulations \citep{ost_steiner+al2010}.

A more comprehensive view of vortices in the solar atmosphere reveals that there exists
not only vertical and horizontal vortices (which are probably easiest to detect)
but an unsteady network of highly tangled filaments of swirls, some of which 
protruding above the optical surface, for example in the form of arc shaped
filaments \citep{ost_moll+al2011}.

Attempts to measure the angular distribution of the magnetic field
in quiet regions and internetwork regions of the Sun 
\citep{ost_lites+al2007,ost_lites+al2008,ost_orozco-suarez+al2007,%
ost_martinez-gonzalez+al2008,ost_beck+rezaei2009,ost_asensio-ramos2009,%
ost_danilovic+al2010,ost_stenflo2010,ost_ishikawa+tsuneta2011,ost_borrero+kobel2012} 
have led to controversial results. Basic difficulties arise 
from the fact that in such regions, (i) the
Zeeman splitting of spectral lines is much smaller than the line width because
the field strength is mostly weak or spatially not resolved (small filling factor)
so that the field strength cannot be determined by simply measuring the Zeeman 
splitting. Instead,  detailed radiation transfer and atmospheric modeling is required
for correctly interprete polarimetric measurements. Even more problematic
for the determination of the inclination angle is (ii) that the linear polarization
depends in the weak field limit quadratically on the field strength while the circular 
polarization has a linear dependency. Since both polarization states are subject to 
a common noise level, the transversal field component is more severely affected by 
noise than the longitudinal component. Monte Carlo simulations with different noise
levels added to synthesized Stokes profiles \citep{ost_borrero+kobel2011} show, that
a Stokes-inversion procedure may interpret pure noise as transversal fields with 
a strength of, e.g., 80~G for a noise level of $\sigma = 1.0\times 10^{-3}\,I_c$.
We propose a less biased determination of the mean horizontal to vertical field
strength by applying a common noise criterion to the field strengths in the real 
physical domain instead of to signals in the Stokes space.

Establishing a bivariate scatter plot of the blue amplitudes of Stokes $V$ of
\ion{Fe}{i} $\lambda\lambda$ 630.15~nm and 630.25~nm for each pixel of a quiet Sun
region, reveals two populations of points \citep{ost_stenflo2010}: (i) a population
with a slope that is compatible with the weak field limit and (ii) a population
that seems to be only compatible with kG fields. We have analyzed synthesized 
Stokes-$V$ profiles from
an MHD-simulation with the result that they too show the same two populations in 
corresponding scatter plots but cannot be explained in terms of intrinsically 
strong and weak fields. Rather, we find both populations stemming from intrinsically 
weak fields but from different locations. The population compatible with the
weak field limit typically stems form clusters of pixels located within granules. 
The population that was formerly
interpreted as being due to intrinsically strong field of small filling factors
stems from pixels located at the boundaries between granules and the inter-granular 
space which harbor (weak) canopy fields with a field strength that abruptly increase with
height in the atmosphere and show a downdraft in the deep photospheric layers 
(below the canopy field). So far, we have not analyzed the details of the radiation 
transfer that creates under these circumstances Stokes-$V$ profiles that
behave, regarding their amplitude ratio, like stemming from kG-fields but in reality 
originate from weak fields. The two lines
\ion{Fe}{i} $\lambda\lambda$ 630.15~nm and 630.25~nm do not form under the same
conditions, in particular not at the same height in the atmosphere. Therefore,
it can be either that the two lines sample different field strengths because of the
canopy character of the field, which implies steep field gradients, or that they
are differently affected by the downflows in the deep layers, creating
different Stokes-$V$ asymmetries, or that thermal effects play a role.
From the present analysis we conclude that the two populations are 
\emph{not} due to a dichotomy between ``collapsed'' and ``uncollapsed'' 
fields but due to a distinct difference in the formation of profiles and in the 
physical conditions at the location of the formation, i.e., the granule interiors 
and at the boundaries of granules, where canopy fields prevail. It is very 
likely, that this explanation also holds for the two populations of actually
observed Stokes-$V$ profiles.

Finally, we show, that a new simulation in which convective updrafts across the 
lower boundary advect horizontal fields of 50~G strength into the
computational domain, yield a dominance of the horizontal vs.\ the vertical field 
throughout the photosphere, increasing with height. This is in agreement with and 
corroborates previous simulation results by
\citet{ost_schuessler+voegler2008}, \citet{ost_steiner+al2008}, and 
\citet{ost_danilovic+al2010}.

\acknowledgements The authors acknowledge insightful discussions on small-scale
filament formation within the ISSI International Team lead by Irina N.~Kitiashvili 
at ISSI (International Space Science Institute) in Bern. We are grateful to
Jan O.~Stenflo for letting us use his IDL program for establishing line-ratio
scatter plots and to Juan Manuel Borrero for various comments on the manuscript.

\bibliography{steiner}

\end{document}